\documentclass[preprint,12pt]{aastex}
\usepackage{rotating}
\shorttitle{Galactic Extinction Properties}

\begin{document}

\title{Ultraviolet Extinction Properties in the Milky Way}
\author{Lynne A.\ Valencic\altaffilmark{1,2},Geoffrey C.\ 
Clayton\altaffilmark{1}, \& Karl D. Gordon\altaffilmark{3}}
\altaffiltext{1}{Department of Physics \& Astronomy, Louisiana State
University, Baton Rouge, LA 70803; valencic@phys.lsu.edu; gclayton@fenway.
phys.lsu.edu}
\altaffiltext{2}{Currently at Instituto de Astronomi\'{a}, UNAM, Apartado 
Postal 877, Ensenada, Baja California, 22800, Mexico}
\altaffiltext{3}{Steward Observatory, University of Arizona, Tucson, AZ
85721; kgordon@as.arizona.edu}


\begin{abstract}

We have assembled a homogeneous database of 417 ultraviolet (UV) 
extinction curves for reddened sightlines 
having International Ultraviolet Explorer (IUE) spectra. We have 
combined these with optical and 2MASS 
photometry allowing estimates of the ratio of total-to-selective extinction,
R$_V$, for the entire sample.  Fitzpatrick-Massa (FM) parameters 
have also been found for the entire sample. 
This is the largest study of parameterized UV extinction curves  
yet published and it covers a wide range of environments, from 
dense molecular clouds to the diffuse interstellar medium (ISM), with 
extinctions A$_{V}$ ranging from 0.50 to 4.80.  It is 
the first to extend far beyond the solar neighborhood and into 
the Galaxy at large, with 30
sightlines having distances $>$ 5 kpc.  Previously, the 
longest sightlines with FM parameters and R$_{V}$ extended $\sim$ 1 kpc. 
We find that (1.) the CCM extinction law applies for 93\% of the 
sightlines, implying that dust processing in the Galaxy is efficient 
and systematic; (2.) the central wavelength of the 2175 \AA\ bump is constant; 
(3.) the 2175 \AA\ bump width is dependent on environment. 
Only four sightlines show systematic deviations from CCM, HD 29647, 62542, 
204827, and 210121. These sightlines all sample dense, molecule-rich clouds.
The new extinction curves and values of R$_{V}$ allow us to revise the 
CCM law. 

\end{abstract}


\section{Introduction}

Astronomical ultraviolet studies were not possible before the 
advent of rocket-borne instruments and satellites due to 
atmospheric absorption and Rayleigh scattering.  The first 
satellite capable of UV observations, OAO-2, was launched 
in 1968, and was followed by a host of other satellites
(Cox 2000), including Thor Delta-1 (TD-1), Astronomy Netherlands
Satellite (ANS), and IUE, among others.  
The data obtained by these instruments are responsible for 
the vast majority of contributions to studies of interstellar 
dust, as dust grains preferentially extinguish 
short-wavelength light.  By comparing heavily extinguished stars 
against their intrinsic fluxes (the pair method), we can better 
understand grain properties and composition.  This is important 
for two reasons.  
First, solid knowledge of grain properties will allow us to build
more realistic grain models and correct for the effects of dust
in stellar and galactic photometry and spectra more accurately.
Second, dust grains are extremely sensitive to their environment,
and can yield much information about local conditions and the
physical and chemical processes which affect grains. 

The pair method, first used by Bless \& Savage 1970, has 
since been used by many others to make extinction curves and 
study the ISM in the Milky Way (e.g., York et al. 1973; Nandy et 
al. 1975, 1976; Koorneef 1978; Seaton 1979; Snow \& Seab 1980; 
Kester 1981; Meyer \& Savage 1981; Aiello et al. 1982; 
Massa, Savage \& Fitzpatrick 1983; Massa \& Savage 1984;
Savage et al. 1985; Fitzpatrick \& Massa 1986, 1988, 1990, hereafter
FM86, FM88, FM90; Clayton \& Fitzpatrick 1987; Aiello et al. 1988;
Cardelli et al. 1988, 1989, hereafter CCM; Papaj, Krelowski, \& 
Wegner 1991;  
Papaj \& Krelowski 1992, and Jenniskens \& Greenberg 1993, hereafter 
JG93) as well as in other
galaxies (Borgman et al. 1975; Nandy et al. 1981; Koorneef
\& Code 1981; Clayton \& Martin 1985; Fitzpatrick 1985, 1986;
Clayton et al. 1996; Bianchi et al. 1996; Gordon \& Clayton 1998;
Misselt et al. 1999; Gordon et al. 2003). 

There are significant differences between dust grain properties 
in the Galaxy and those found in the Small and Large Magellanic 
Clouds (SMC and LMC, respectively). The SMC Bar has very strong, 
linear FUV extinction, while the sightlines toward the LMC2 Supershell 
and the average LMC curve have a weaker bump and stronger FUV 
extinction than the average Galactic R$_V$=3.1 curve.  There are 
also significant
differences in metallicities and the gas-to-dust ratios 
between the Galaxy, LMC, and SMC (Bohlin et al. 1978; 
Gordon et al. 2003; Luck \& Lambert 1992).

A major leap forward in understanding interstellar extinction
occurred in 1989, when CCM suggested a mean extinction relation 
that depended on one parameter, the ratio of total-to-selective 
extinction, R$_V$ (=A$_{V}$/E(B-V)).  With only very few exceptions, 
Galactic extinction curves tend to follow this law within the 
uncertainties of the calculated R$_V$ values and the extinction 
curves (Fitzpatrick 1999; Clayton et al. 2000).
The Galactic diffuse ISM is well described by a curve where
R$_V$ = 3.1.  R$_V$ can also be thought of as a rough indicator
of grain size, with low R$_V$ sightlines having more small grains
than high R$_V$ sightlines.
The CCM extinction relation generally does not
apply outside the Galaxy, although recent work has shown that
there are sightlines toward the LMC which may follow CCM (Gordon et
al. 2003).  This law essentially replaced the Galactic average extinction
curve (Seaton 1979) with a family of R$_V$-dependent extinction
curves, with each curve representing a mean curve for sightlines
of that particular value of R$_V$.  This also showed that many of
the so-called ``anomalous'' sightlines (e.g. Clayton \& Fitzpatrick
1987) were in fact normal.  In addition to finding 
this relationship, CCM pointed out the
usefulness of normalizing extinction curves to A$_{V}$, rather than
E(B-V), as was usually done.  Compared to E(B-V),
A$_{V}$ is a more basic quantity, and it is a direct measure of 
optical depth along a line of sight. 
Galactic extinction curves 
can be fit with six parameters (Fitzpatrick \& Massa 1990) which 
allow for quantitative analysis of extinction characteristics. 

Despite all the work with extinction curves that has been
done, a homogeneous database of
Milky Way extinction curves does not exist.  The many previous studies, 
listed above, used data from different instruments, and variations of 
the pair method to construct extinction curves.  So, the results 
of these studies cannot be compared easily.  Also, these sightlines 
make up only about half of reddened sightlines available in the IUE Archive.    
To illustrate the disparateness of previous works, Table \ref{tab:previous_work}
compares various aspects of previous studies to the present work.
Column (1) lists authorship, (2) lists the number of sightlines, 
(3) lists the instrument the data was from, (4) shows if  
FM parameters were found, (5) shows whether the extinction curves were 
constructed using the MK classification or UV stellar features,
and (6) shows if IR photometry, specifically R$_V$, was
considered for the entire sample.

In order to solve this problem, we have built a homogeneous database
of over four hundred extinction curves constructed in a uniform manner.
Over 150 of these sightlines extend beyond 2 kpc, and 30 extend beyond 
5 kpc, thus sampling a much larger volume of the Galaxy than ever before.  
This is a much larger and more complete database than any done
previously. 
Until the recent release of the 2MASS 
database, JHK photometry was available for only a small fraction of the 
stars in the sample.  Now, estimates of R$_V$ are available for the
entire sample. Thus, it is now possible to draw a more coherent 
picture of extinction in the Galaxy.
In the following sections, we describe our database of extinction 
properties, present the results of our analyses, and discuss 
the implications for dust grains. 

\section{UV Data and Extinction Curves}

For the sake of homogeneity, Hiltner's UBV photometry
(Hiltner 1956; Hiltner \& Johnson 1956) was preferred, when possible,
and was obtained for 192 sightlines.
In his massive compilation and assessment,
Nicolet (1978) found that Hiltner's photometry was consistently of
high quality.  Systematic errors were typically less than 0.01 mag
(Hiltner 1956) and photometry agreed well with the works of others
(Nicolet 1978).  
Photometry in the JHK bands was available for all the stars in the sample 
from the 2MASS database (Cutri et al. 2003).

Reddened stars, observed by IUE, having spectral types O3 to B5 were
selected for the database. An exception is HD 29647, a B8 III
star, included because it has been well studied in the past (e.g., 
Cardelli \& Savage 1988).
This range of spectral types was chosen to minimize the effects
of spectral mismatch in their extinction curves (Massa et al. 1983). 
Thorough discussions of the uncertainties in the pair method can be found 
in Massa et al. (1983), Savage et al. (1985), Aiello et al. 
(1988), and Gordon \& Clayton (1998).  Only ``normal'' stars were included 
which had good unreddened
UV comparison stars.  
The lower limit on E(B-V) was 0.20.

IUE spectra for all of the stars in the sample 
were obtained from the Multimission
Archive at Space Telescope (MAST). 
The archive spectra were reduced using NEWSIPS and then 
recalibrated using the method developed by Massa \& Fitzpatrick (2000). 
The signal-to-noise of the NEWSIPS IUE spectra have been improved 
by 10-50\% over those of IUESIPS IUE spectra (Nichols \& Linsky 1996). 
Low dispersion LWR/LWP and SWP spectra were selected, 
from either aperture. Multiple spectra from one camera were averaged 
and then the long and short-wavelength segments were merged at the 
shortest wavelength of the SWP. 
The wavelength coverage is $\sim$1200 -- 3200 \AA\ with a resolution 
of $\sim$5 \AA.
  
The standard pair method, in which a reddened star is compared with an 
unreddened one of the same spectral type, was used to construct each 
sightline's extinction curve (Massa et al. 1983).  The comparison 
stars were selected from Cardelli, Sembach, \& Mathis (1992) and 
dereddened.  The spectral matches were made on the basis of comparing 
the UV spectra
of pairs of stars rather than matching their visible spectral types. 
R$_V$ was estimated from the JHK colors as described in
Fitzpatrick (1999).  A$_{V}$ was found using R$_V$ and E(B-V).
In general, our UV spectral
classifications are in good agreement with those which have been 
published previously (e.g., 
Aiello et al. 1988; Clayton \& Fitzpatrick 1987; Papaj et al. 1991). 
Photometry, sources, MK and UV spectral types, reddenings, and 
calculated R$_V$ values for all sightlines can be found 
in Tables \ref{tab:phot}, \ref{tab:citation_key}, and \ref{tab:ext_query}. 

The resulting extinction curves were fit with the Fitzpatrick-Massa (FM) 
parameterization (Fitzpatrick \& Massa 1990) and then normalized to A$_V$.  
Setting $x=1/\lambda$, the FM fitting function is given by 

\begin{equation}
  \label{eq:FMfit}
  k(x) = \frac{E(\lambda-V)}{E(B-V)} = c_1 + c_{2}x + c_{3}D(x,\gamma,x_{0}) + c_{4}F(x) \\
\end{equation}

where 

\begin{equation}
  \label{eq:drude}
  D(x,\gamma,x_{0}) = \frac{x^{2}}{(x^{2}-x_{0}^{2})^{2}+(x \gamma)^{2}} \\
\end{equation}

and for $x \ge 5.9~ \mu$m$^{-1}$, 

\begin{equation} 
  \label{eq:nonlinear}
  F(x) = 0.5392(x-5.9)^{2}+0.05644(x-5.9)^{3} \\
\end{equation}

while $F(x)=0$ for $x < 5.9 ~ \mu$m$^{-1}$. 

Eqn. \ref{eq:FMfit} results in 6 parameters, c$_{1}$, c$_{2}$, c$_{3}$, c$_{4}$, 
x$_{0}$, and $\gamma$,  each of which describe different attributes of the curves. 
The first two, c$_{1}$ and c$_{2}$, account for
the intercept and slope of the linear background.  They are not
independent of each other and can be merged into one parameter
(Carnochan 1986, FM88), though that is not done here.  The quantities
c$_{3}$ and c$_{4}$ correspond to the strength of the bump and the
curvature of the FUV rise.  The last two parameters,
$x_{0}$ and $\gamma$, correspond to the central wavenumber and
width of the bump, respectively; $x_{0}$ does not vary greatly from
sightline to sightline, so it may be possible to reduce the
number of parameters to four overall (Fitzpatrick 1999).
As can be seen in Eqns. \ref{eq:drude} and \ref{eq:nonlinear}, the bump is 
fit by a Drude profile and the FUV extinction is fit by a nonlinear function.  
The parameters were found 
using the three-step method of Gordon et al. (2003), over the 
wavelength range 2700 -- 1250 \AA. The FM fit is not reliable 
longward of 2700 \AA~(E. Fitzpatrick 2002, private 
communication). 
The spectra are cut at the blue end at 1250 \AA\ in order to 
exclude the Ly$\alpha$ 
feature at 1215 \AA.   
The normalization of the FM parameters was converted from E(B-V) 
to A$_{V}$ (JG93):

\begin{eqnarray*} 
  \frac{A(\lambda)}{A(V)}= \frac{k(x)}{R_{V}} + 1.0  \\
\end{eqnarray*}

so that the FM parameters become

\begin{eqnarray*} 
  c_{1}^{AV} = c_{1}/R_{V}+1.0~~~~~~~ \\
  c_j^{AV} = c_{j}/R_{V},~ j=2,3,4
\end{eqnarray*}

The resulting normalized FM parameters are 
listed in Table \ref{tab:FM_query}. The dust environment was assessed  
by considering the traditional density measure, $A_{V}/d$; JG93 found that 
sightlines passing through dense regions had $A_{V}/d >$ 0.9 mag kpc$^{-1}$, 
while more diffuse sightlines had $A_{V}/d <$ 0.9 mag kpc$^{-1}$.  
We have adopted this criterion for this study.  This parameter can 
be inaccurate since the density is averaged over the entire 
sightline (JG93).  Other groups have noted that trends in extinction 
characteristics, especially those associated with the bump, can be linked 
to environmental influences (FM86; Cardelli \& Clayton 1991; JG93).  

\begin{figure}[th]
\begin{center}
\includegraphics[scale=.5]{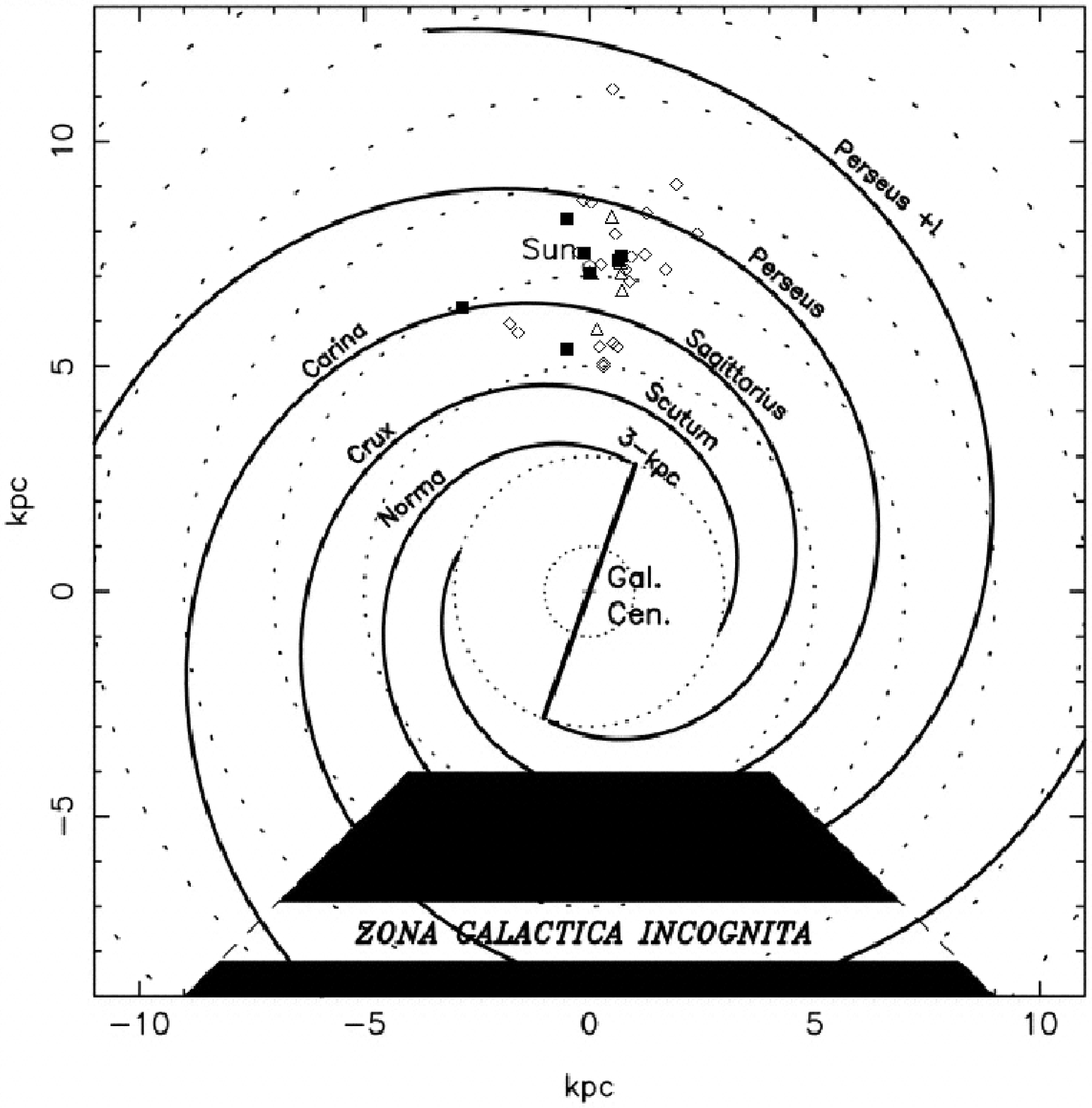}
\includegraphics[scale=.5]{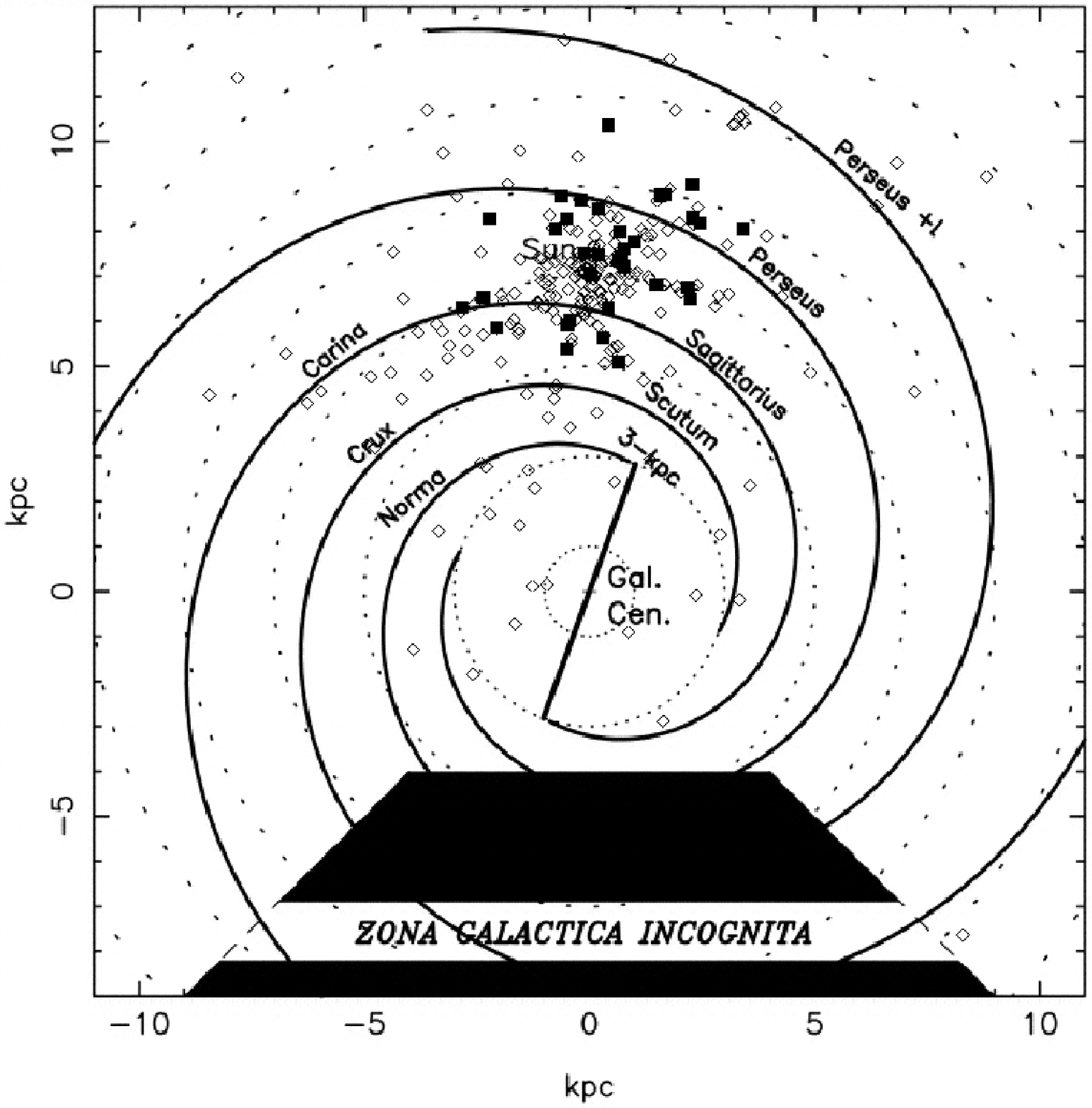}
\end{center}
\caption{Top panel: All sightlines with published FM parameters.  Diamonds: 
Aiello et al. (1988); Triangles: FM90; Squares: OB associations with at least 
one member having FM parameters.  Bottom panel: All sightlines in the 
database.  Diamonds: field stars.  Squares: OB associations with at least
one member having FM parameters. The Galaxy overlay is from Vall\'{e}e (2002). 
\label{fig:galaxy_FMfits}}
\end{figure}

Figure \ref{fig:galaxy_FMfits} shows the distribution of
previously published sightlines in the Galaxy with FM parameters
normalized to E(B-V) and the sightlines in this new database.
Because of our much larger sample, we have more
complete coverage of nearby associations and more distant sightlines
than previous works, pushing out further into the Galaxy, and
sampling a wider variety of environments.  
As distances were estimated using spectroscopic parallaxes, the
uncertainties for stars plotted in Fig.\ \ref{fig:galaxy_FMfits} are
large, with an average uncertainty of 50\%.  

\section{Discussion}

\subsection{Continuum Extinction}

The relationship between the various FM parameters were investigated 
as well as their dependence on R$_V$$^{-1}$.  
These are shown in Figs. 2-8. 
For pairs of parameters with significant correlations,
best-fit lines are plotted, using the 
least absolute deviation method as this is less affected by outliers 
than $\chi^{2}$ minimization.  

\begin{figure}[th]
\begin{center}
\includegraphics[scale=.4]{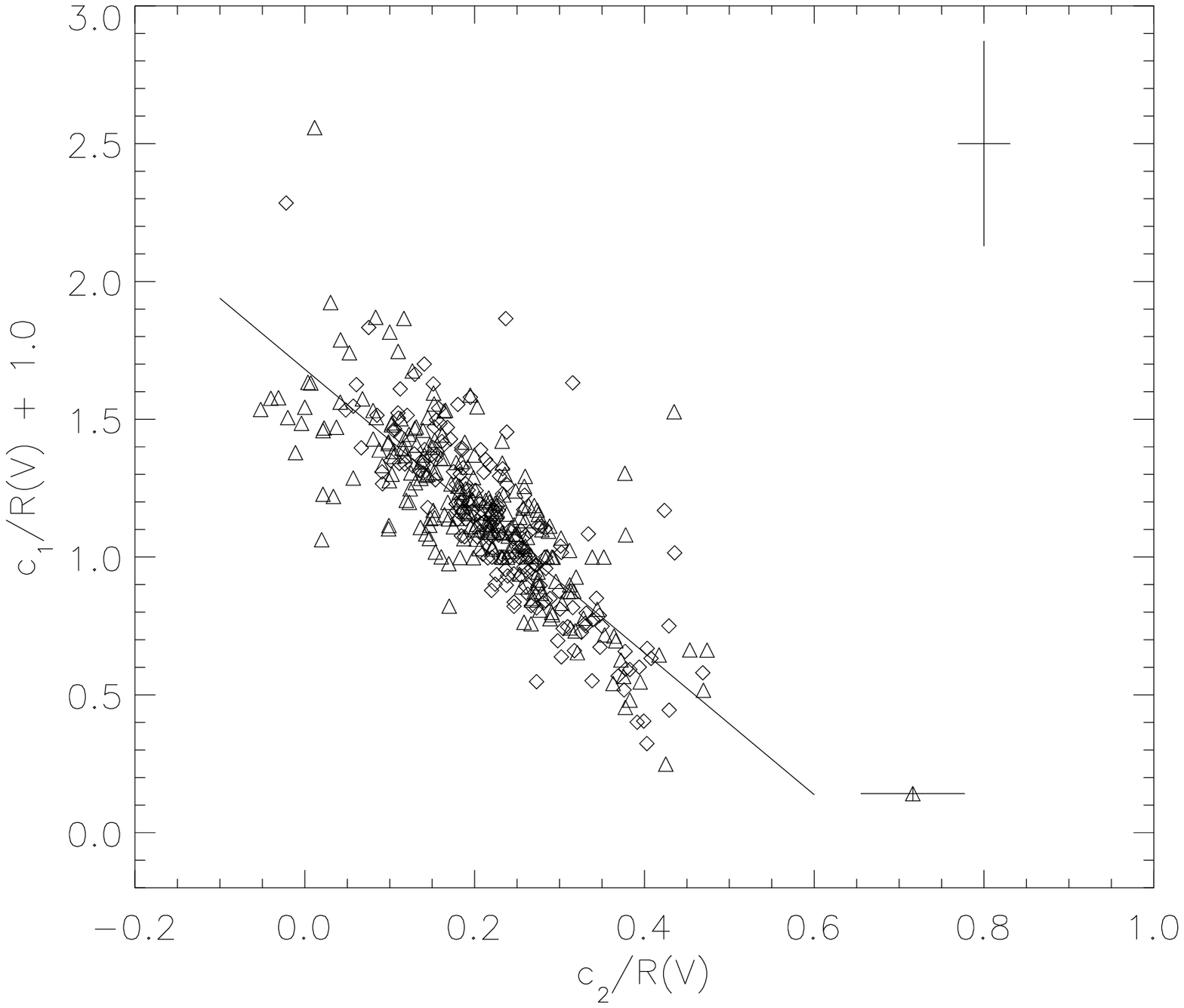}
\includegraphics[scale=.4]{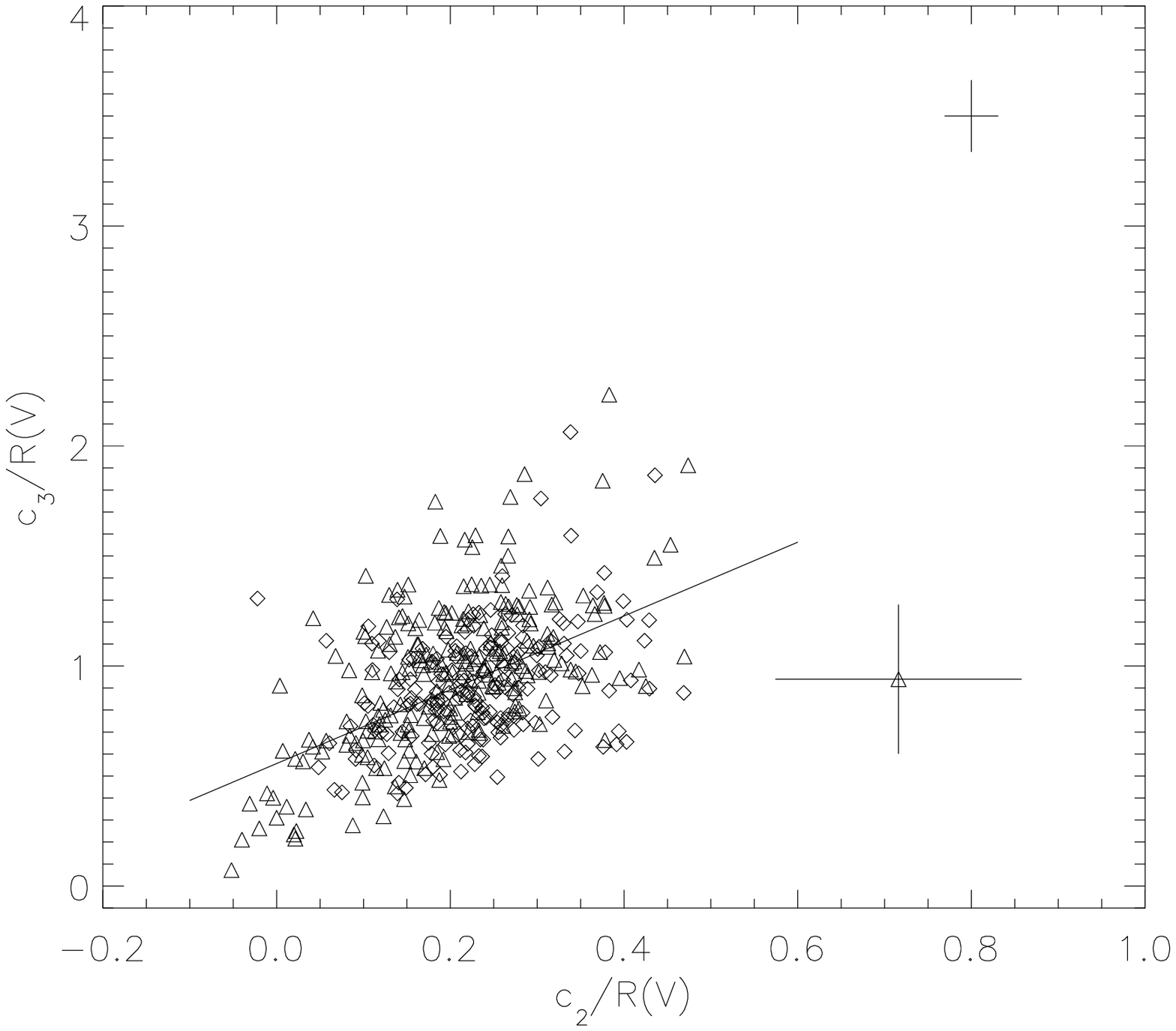}
\includegraphics[scale=.4]{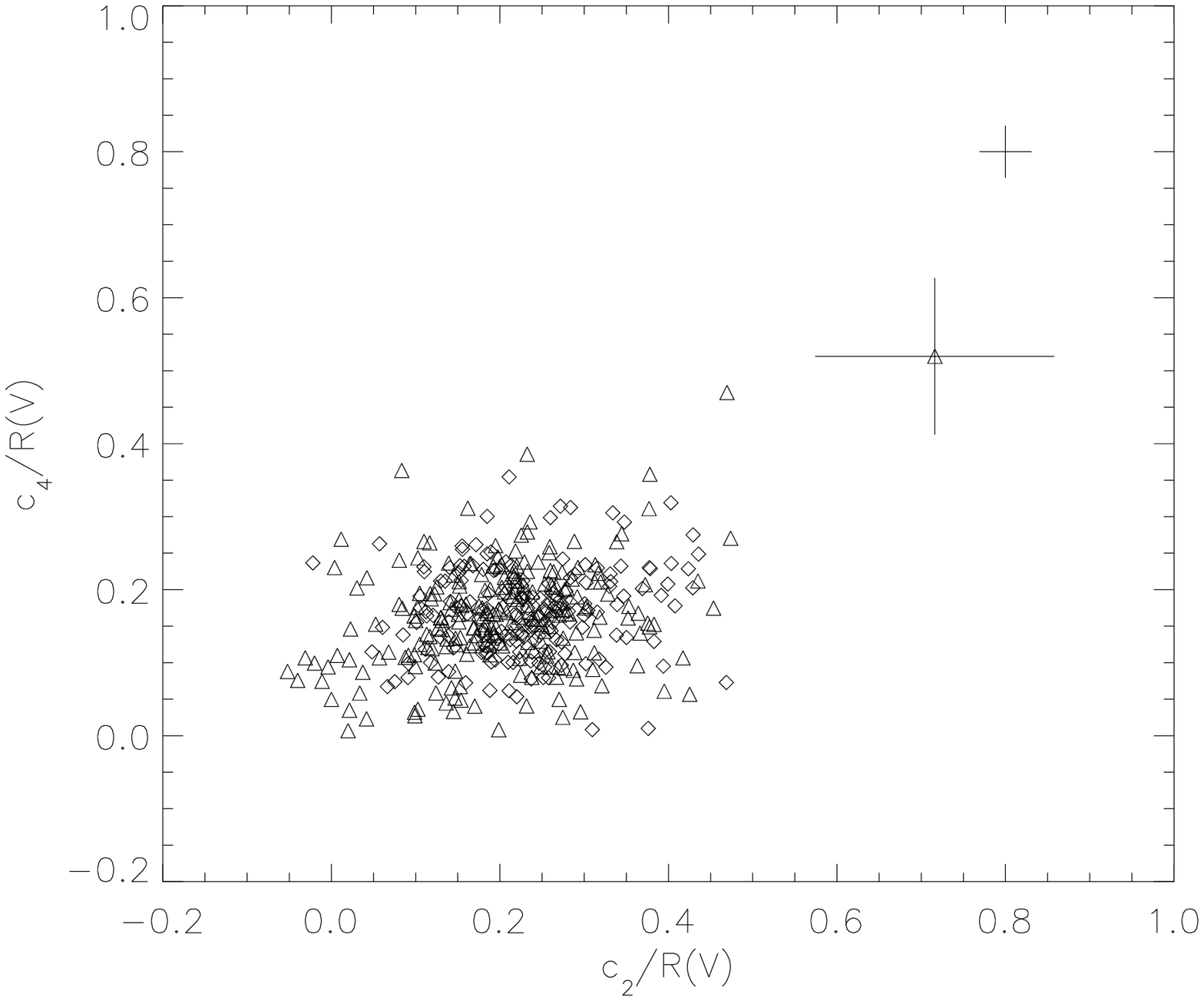}
\includegraphics[scale=.4]{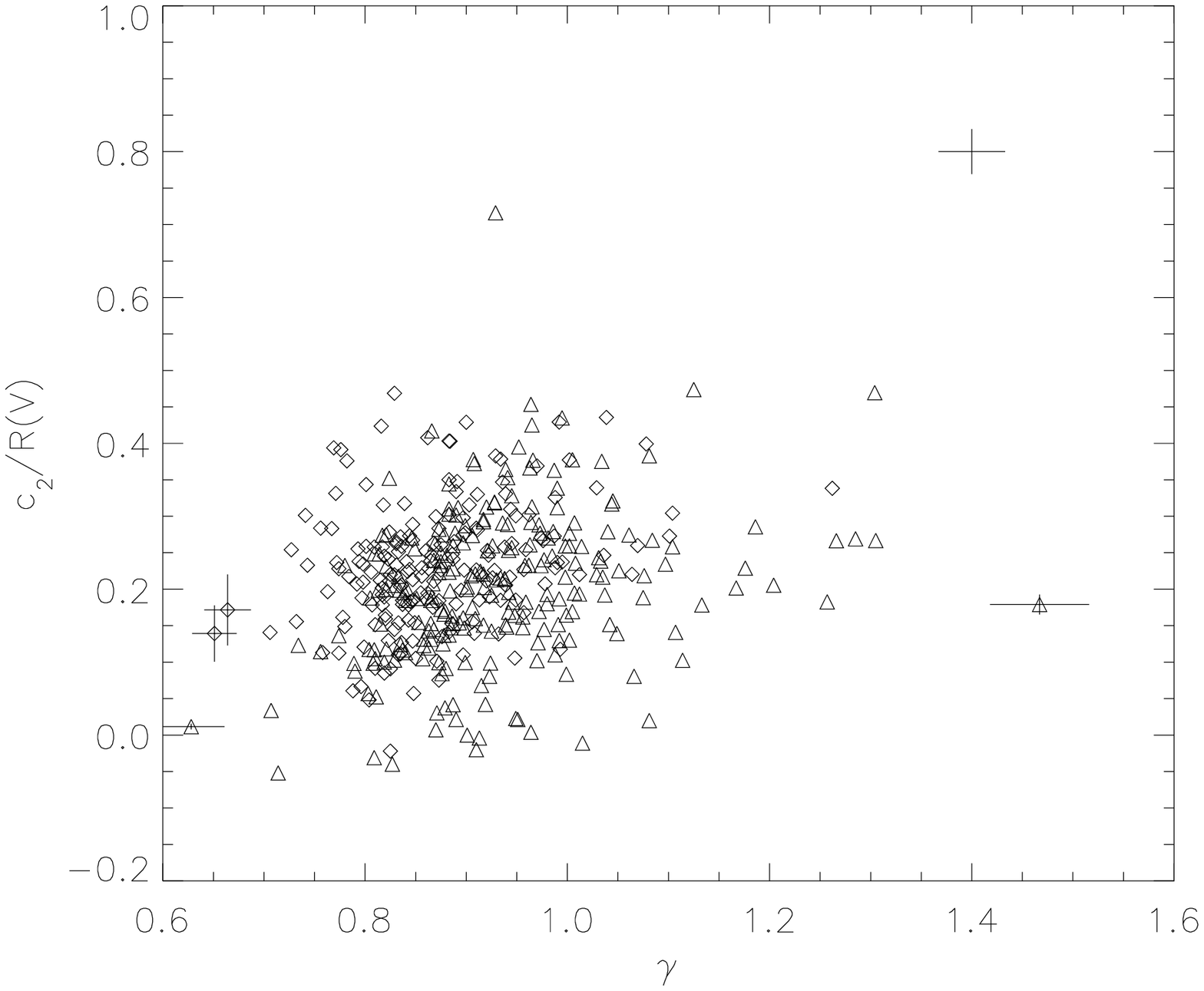}
\includegraphics[scale=.4]{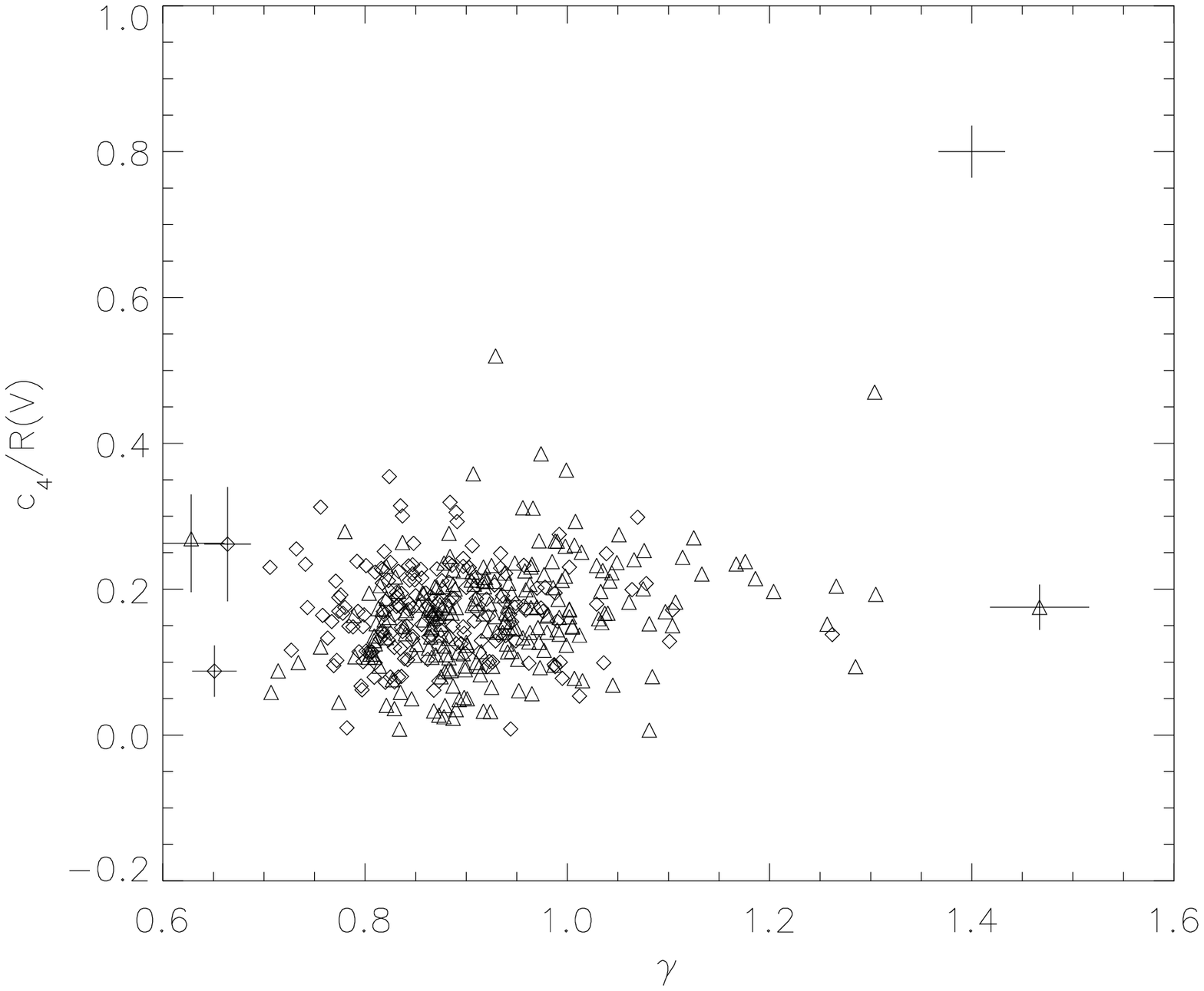}
\end{center}
\caption{FM parameters of database sightlines.  All parameters, except 
x$_{0}$ and $\gamma$, have been divided by R$_V$ and thus normalized to 
A$_{V}$.  Dense sightlines (those with $A_V/d \ge 0.9 mag/kpc$) are indicated by 
triangles, while diffuse sightlines ($A_V/d < 0.9 mag/kpc$) are indicated with 
diamonds.  Representative error bars are indicated. \label{fig:FM_params_cont}} 
\end{figure}

\begin{figure}[th]
\begin{center}
\includegraphics[scale=.7]{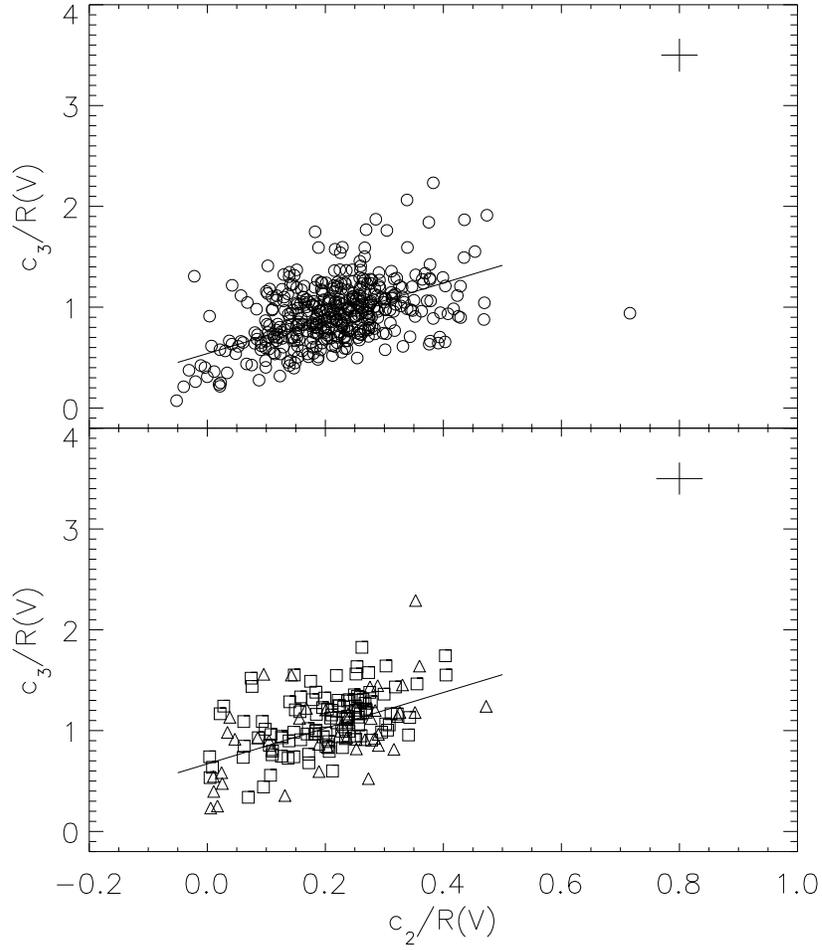}
\end{center}
\caption{Top panel: The FM parameters c$_{2}$ and c$_{3}$ from all 
database sightlines, divided by R$_V$ and thus normalized to A$_{V}$.
Bottom panel: Values of c$_{2}$ and c$_{3}$ from FM88 (triangles) and JG93 
(squares), divided by R$_V$ and thus normalized to A$_{V}$.  Representative 
error bars are indicated in both panels. \label{fig:c2_c3_RVcomp}}
\end{figure}

There is a clear correlation between c$_{1}$/R$_V$ + 1.0 and
c$_{2}$/R$_V$ (see Table \ref{tab:params_correl} and Fig. \ref{fig:FM_params_cont}).
This has been noted before (Carnochan 1986; FM86; JG93).  According to FM88,
c$_{1}$ = -3.00 c$_{2}$ + 2.04, while JG93 find that
c$_{1}$ = (-3.11$\pm$0.11) c$_{2}$ + (2.14$\pm$0.07).
This relation shows how tightly constrained the linear component 
of the extinction is with respect to $x$ (FM88).
There is also a weak correlation between c$_{2}$/R$_V$ and c$_{3}$/R$_V$. 
Neither JG93 nor FM88 found any apparent relationships between the parameters 
c$_{2}$ and c$_{3}$.  Similarly, no correlation was found in our 
sample between these parameters.  
A comparison of the parameters found in this study and those of JG93 and 
FM88 are shown in Fig. \ref{fig:c2_c3_RVcomp}.  The values of 
c$_{2}$ and c$_{3}$ from the other groups have been divided by 
the values of R$_V$ found in this work, and the best-fit line in the 
lower panel was found by considering points from both studies.  It was 
similar to that found here: c$_{3}$/R$_{V}$ = 1.77 c$_{2}$/R$_V$ + 0.67, 
versus c$_{3}$/R$_{V}$ = 1.75 c$_{2}$/R$_V$ + 0.54.  As both 
c$_{2}$/R$_V$ and c$_{3}$/R$_V$ increase with R$_V$$^{-1}$,
and thus are influenced by similar environmental effects, 
perhaps it is not surprising that they
themselves are correlated with each other.

The parameters c$_{4}$/R$_V$ and c$_{2}$/R$_V$ do not show a 
correlation; neither do the parameters c$_{2}$/R$_V$ and $\gamma$. 
These agree with the findings of FM88 and JG93.
There is also no correlation ($r$=0.13) between c$_{4}$/R$_V$ and $\gamma$.  
This is in disagreement with the findings of Carnochan
(1986), FM88 and JG93, the latter of whom found a correlation coefficient
$r$= 0.33 for these parameters.  These groups suggested that 
wider bumps tended to be found along sightlines 
with steep FUV rises, though this correlation was weak.  In the 
present sample, there was no correlation, regardless of whether 
the parameter c$_{4}$ was normalized to E(B-V) or A$_{V}$.
This reinforces the idea that factors influencing the bump 
width are distinct from those influencing the carrier of the 
FUV rise (FM88).

\begin{figure}[th]
\begin{center}
\includegraphics[scale=.35]{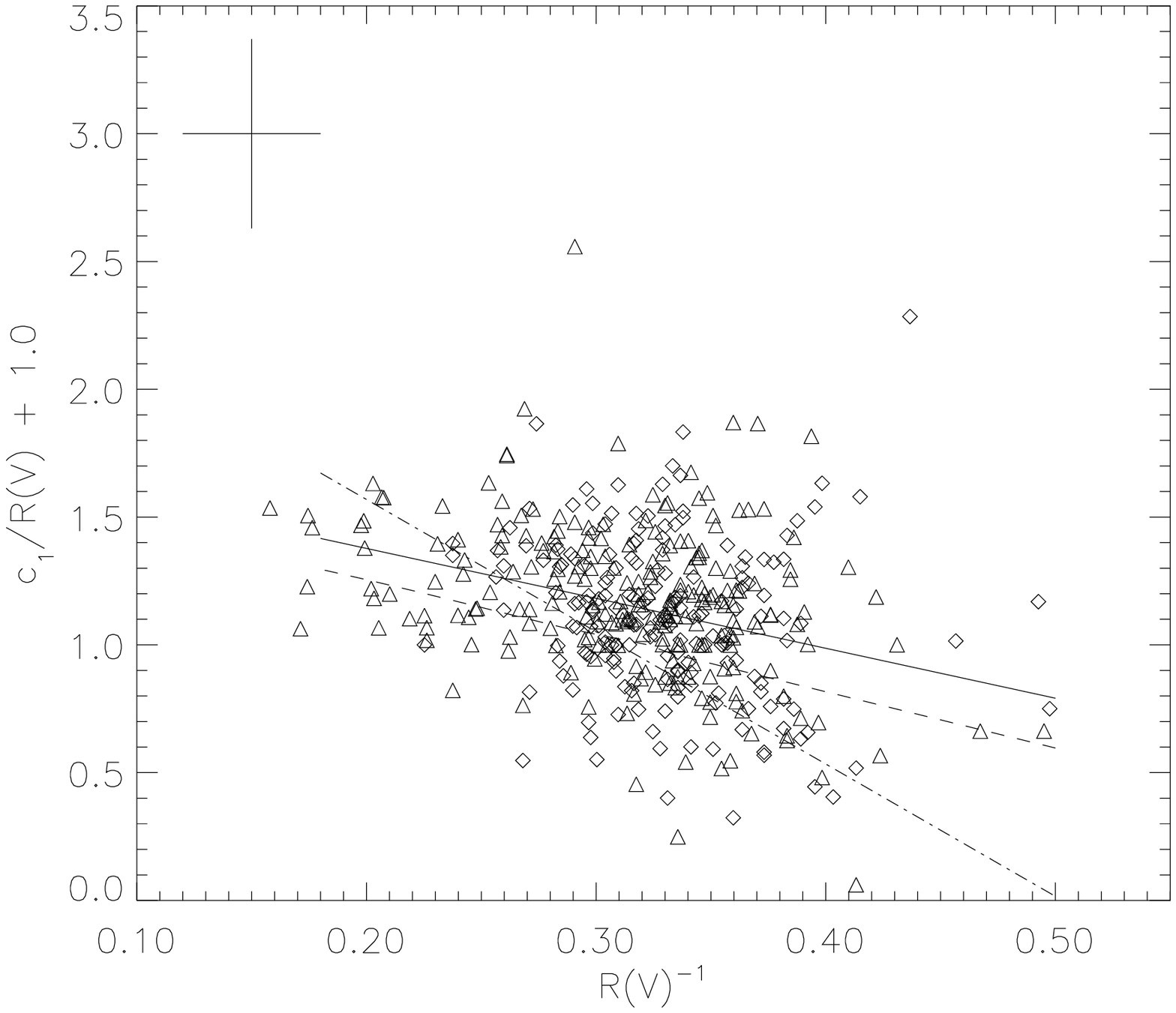}
\includegraphics[scale=.35]{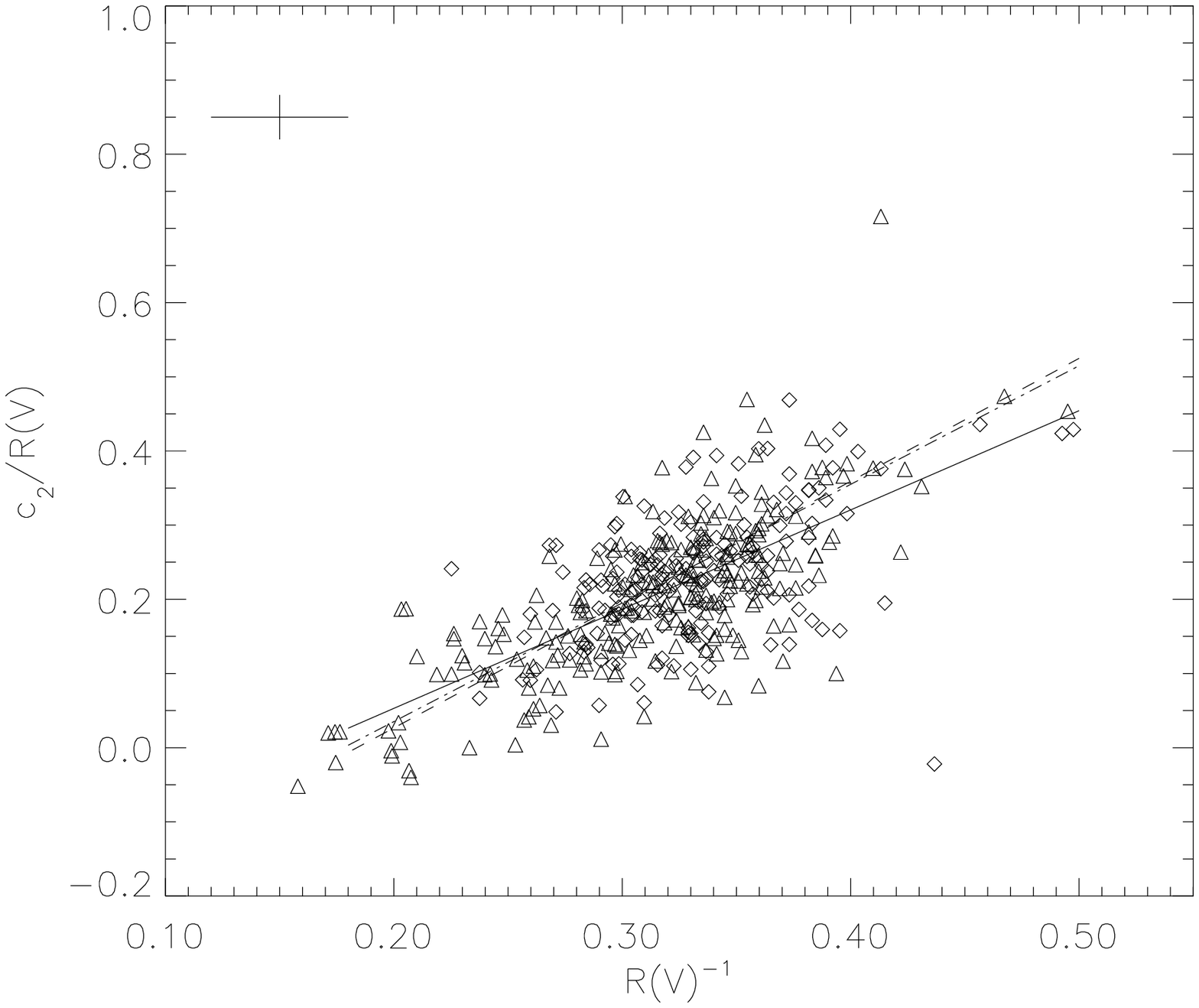}
\includegraphics[scale=.35]{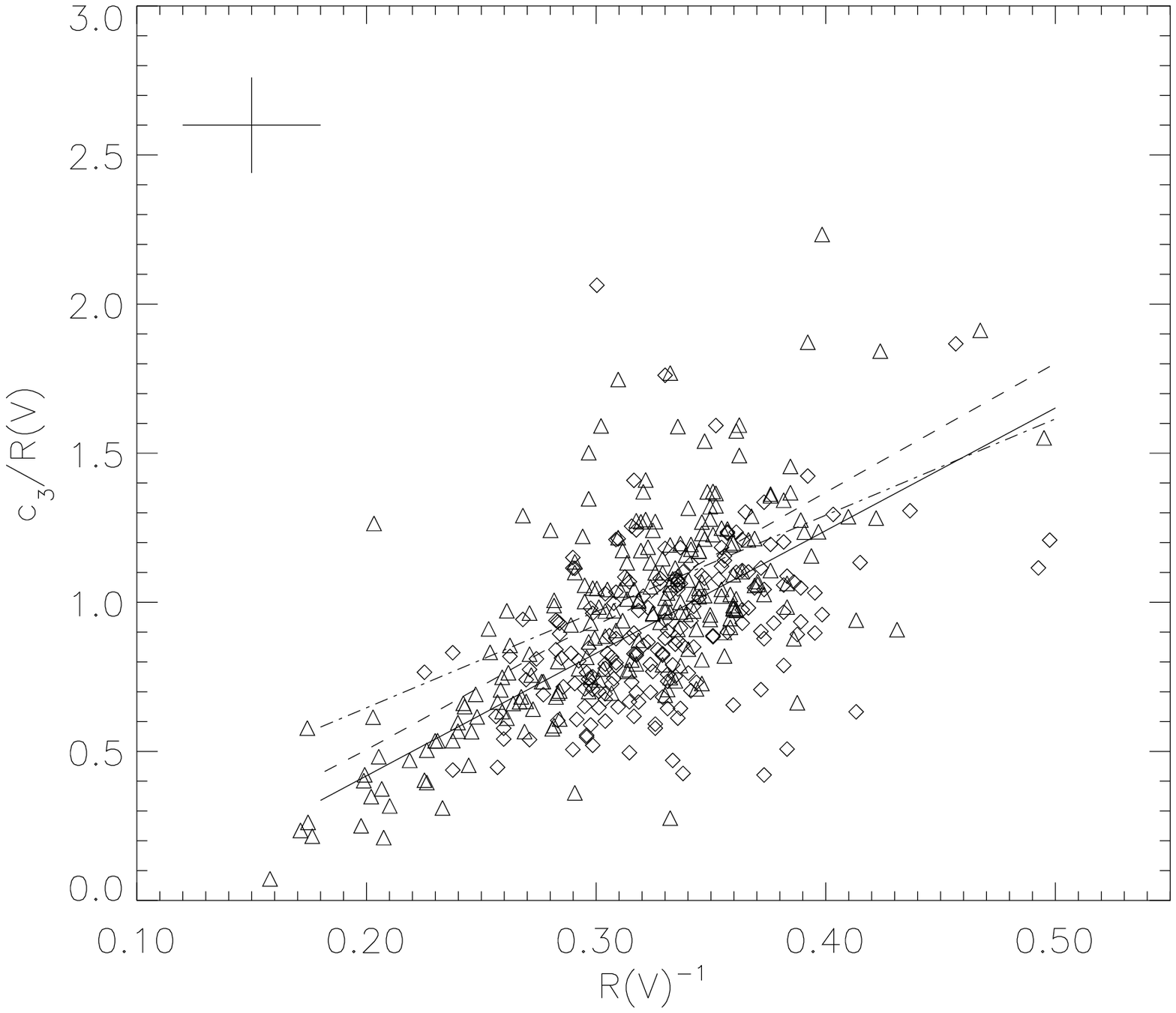}
\includegraphics[scale=.35]{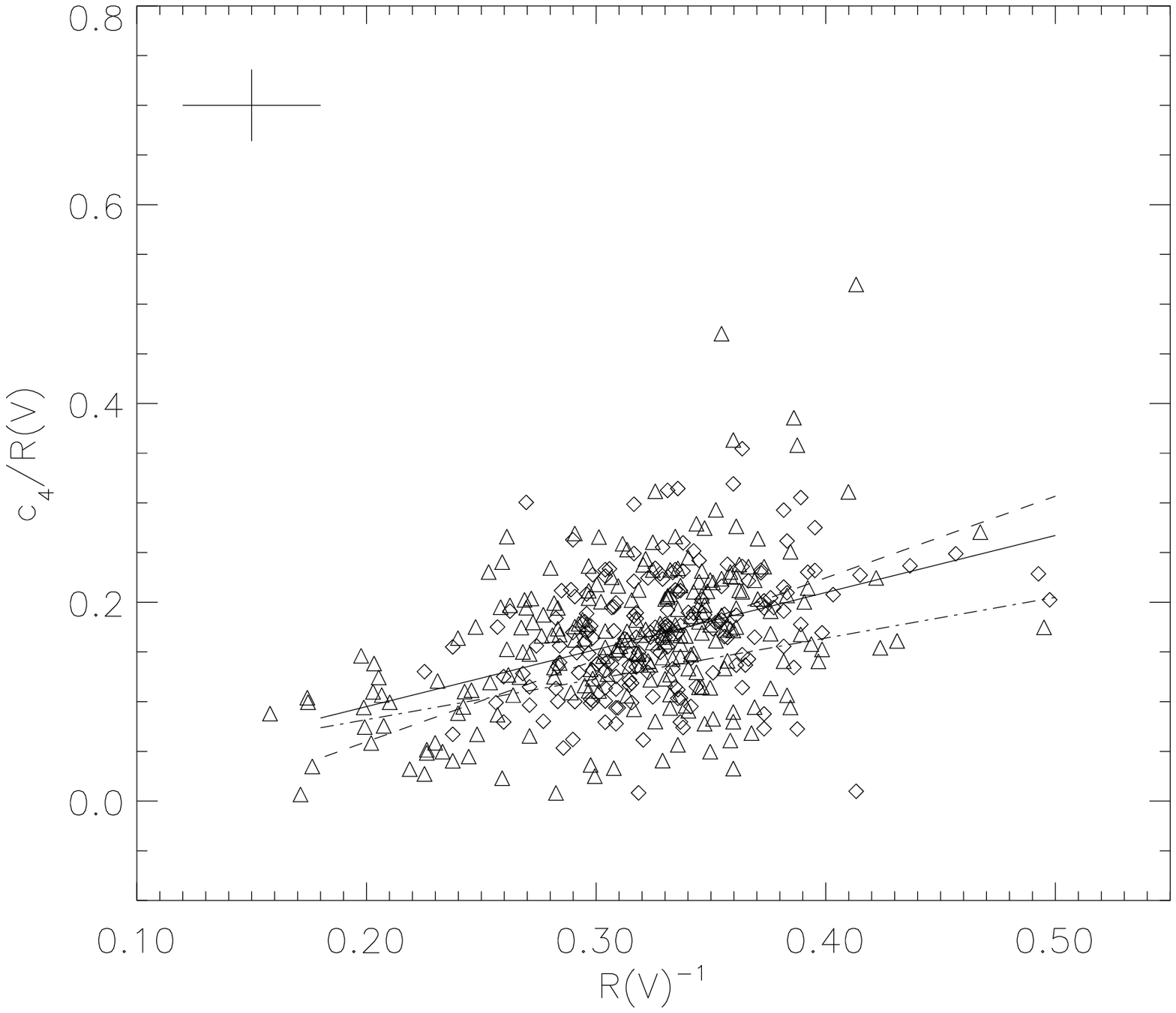}
\includegraphics[scale=.35]{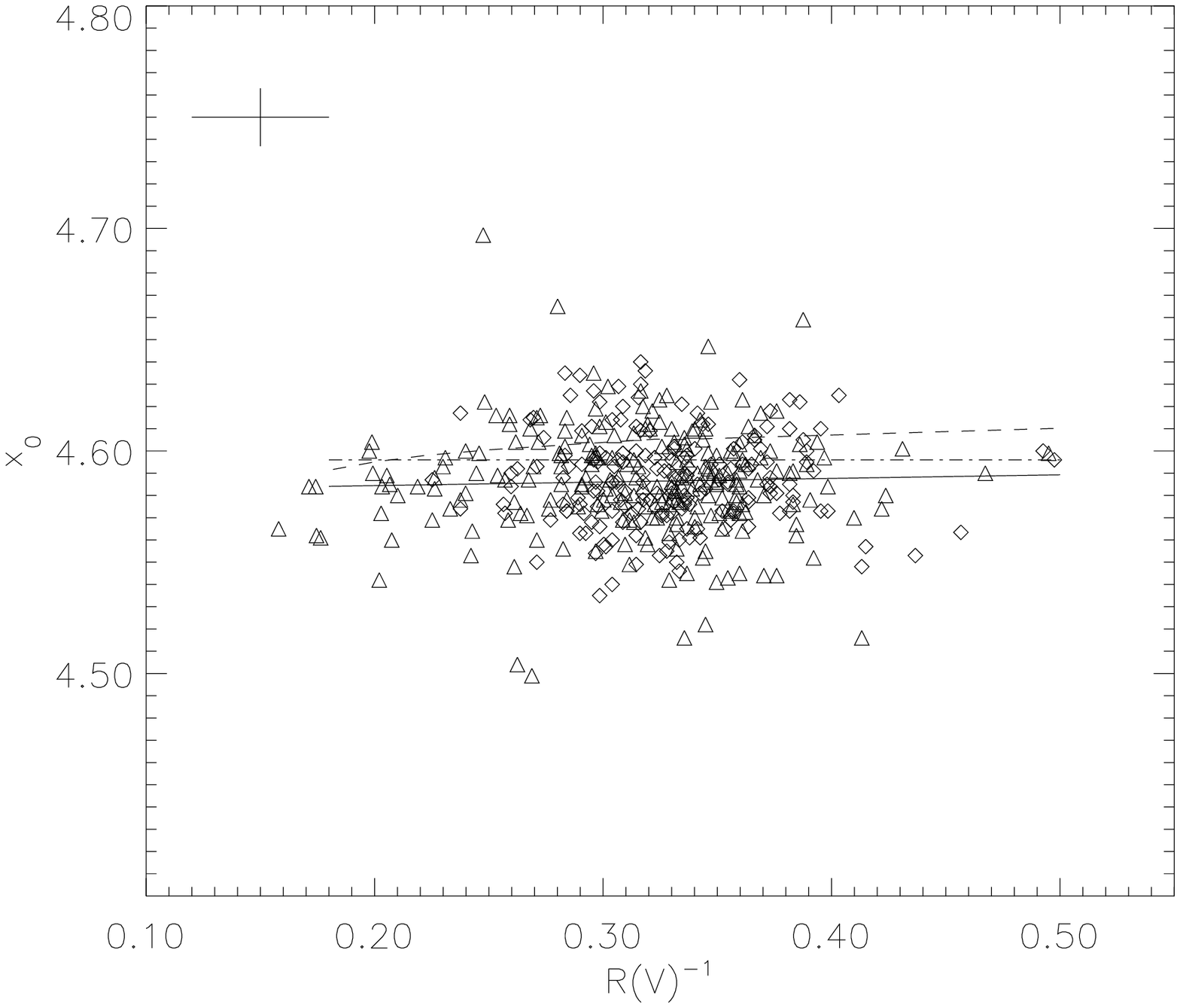}
\includegraphics[scale=.35]{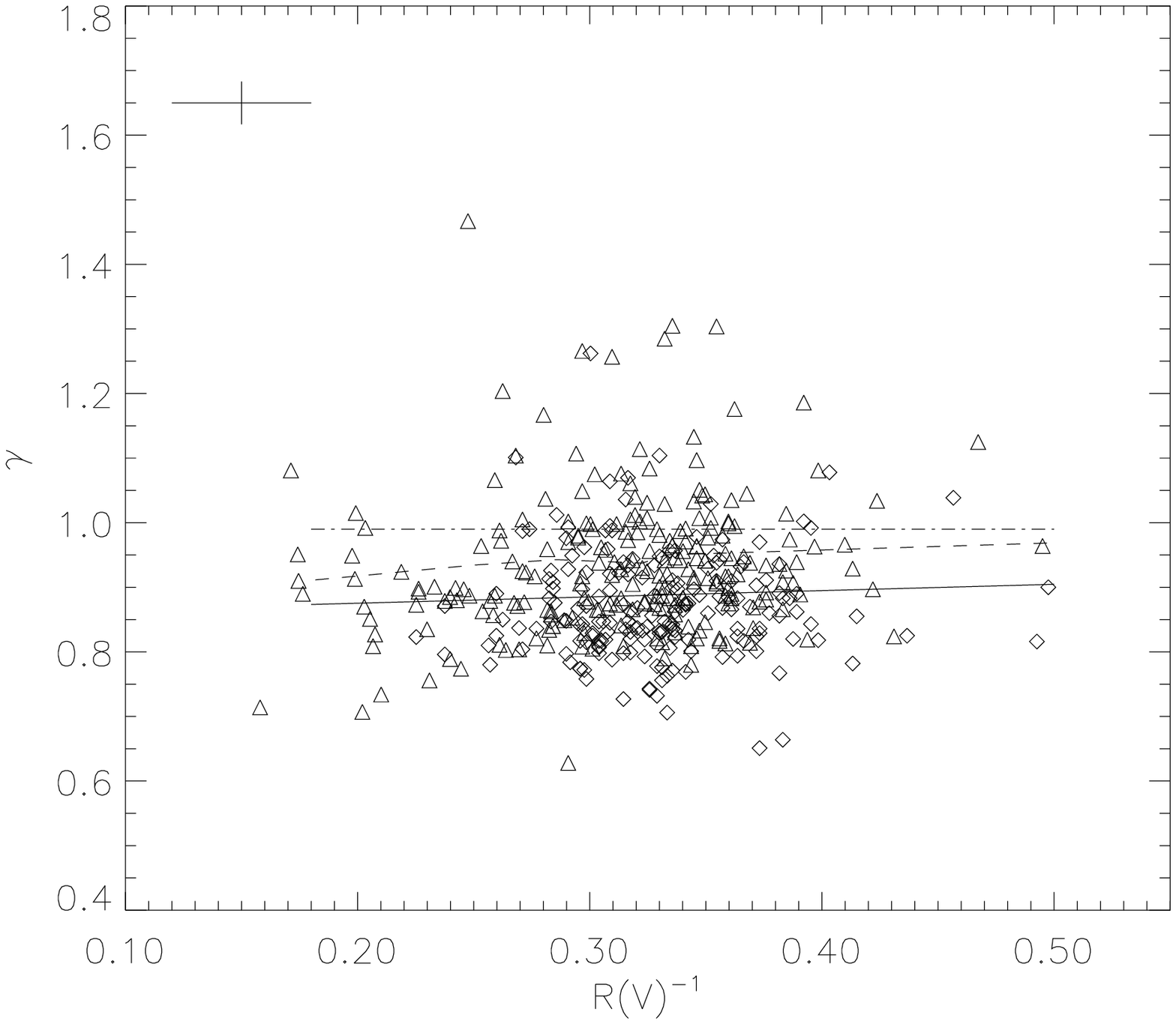}
\end{center}
\caption{FM parameters plotted against R$_V$$^{-1}$.  All parameters, except
x$_{0}$ and $\gamma$, have been divided by R$_V$ and thus normalized to
A$_{V}$.  Symbols are the same as defined in Fig. 2.  The solid lines 
represent the best fits.  The dashed lines indicate the expected values 
from the CCM relation; the dash-dot lines are from Fitzpatrick's (1999) 
reformulation of the CCM law.  Representative error bars are indicated. 
\label{fig:CCM_karl}}
\end{figure}

Correlations
between c$_{2}$, c$_{3}$, c$_{4}$, and R$_V$$^{-1}$, noted by CCM, are
confirmed here, as is the absence of a correlation between x$_{0}$
and R$_V$$^{-1}$.  Also, while $\gamma$ shows real variation, it does
not appear to be linked with R$_V$$^{-1}$.  These can be seen in 
Fig. \ref{fig:CCM_karl}.

\subsection{The 2175 \AA\ Bump} 

\begin{figure}[th]
\begin{center}
\includegraphics[scale=.4]{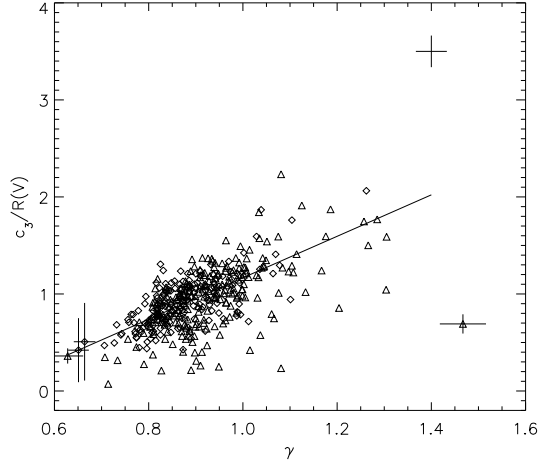}
\end{center}
\caption{Bump height, c$_{3}$, vs. bump width, $\gamma$.
Symbols are the same as defined in Fig. 2.
Representative error bars are indicated. \label{fig:c3_gamma}}
\end{figure}

As the 2175 \AA\ bump is the only spectral feature yet known in the UV, 
parameters which describe it were carefully considered. In Fig. \ref{fig:c3_gamma}, 
a relationship between c$_{3}$/R$_V$ and $\gamma$, very similar to the well-known 
relationship between c$_{3}$ and $\gamma$, can be seen.  The values 
of c$_{3}$/R$_V$ are more confined at lower $\gamma$, but as $\gamma$ 
increases, c$_{3}$/R$_V$ widens its range, with a general trend to
increase with $\gamma$.  This was noted by both FM88 and JG93 in 
their studies of c$_{3}$ and $\gamma$; JG93 attributed it to the fitting 
procedure, but FM88 suggested that these two parameters are truly related 
in some way. 

\begin{figure}[th]
\begin{center}
\includegraphics[scale=.35]{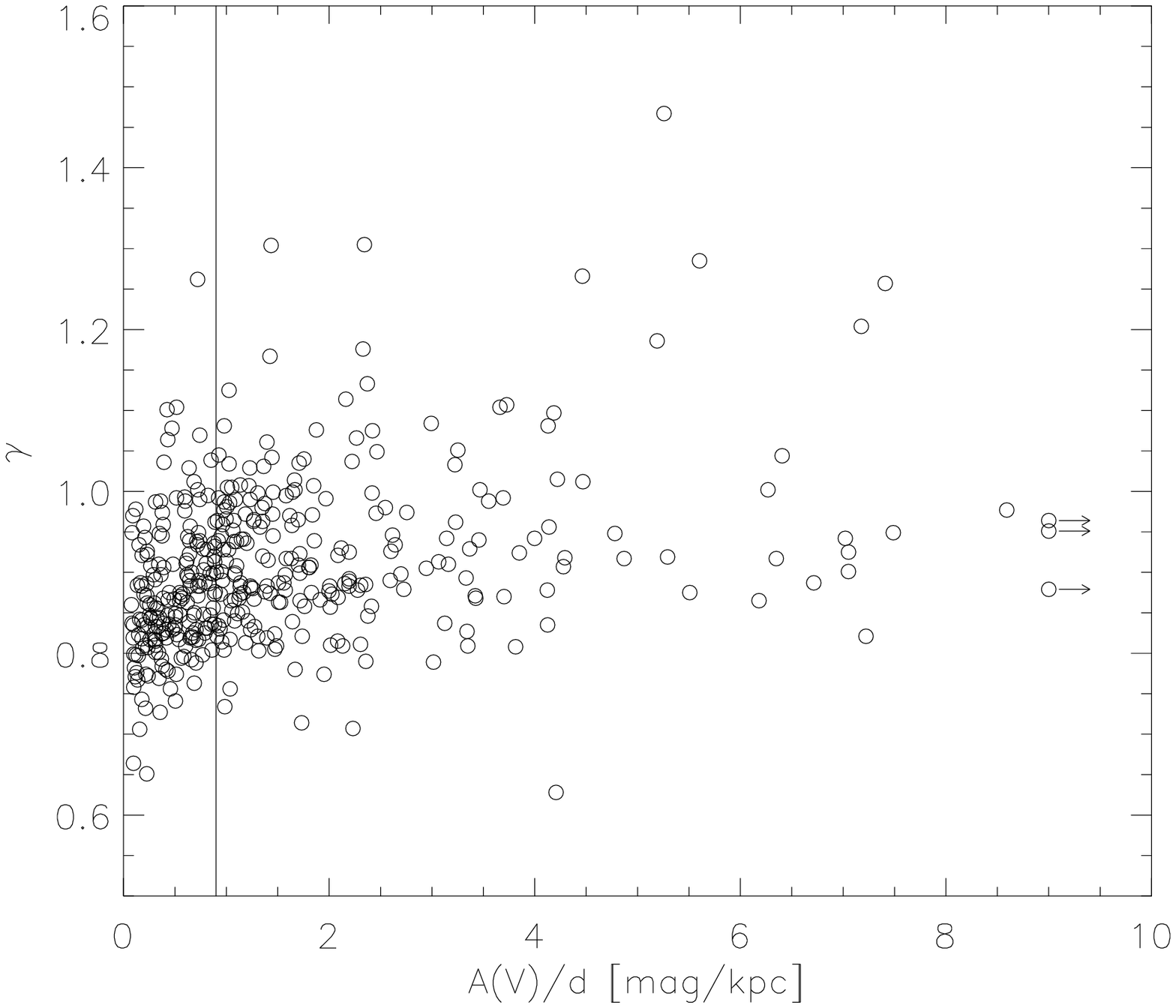}
\includegraphics[scale=.35]{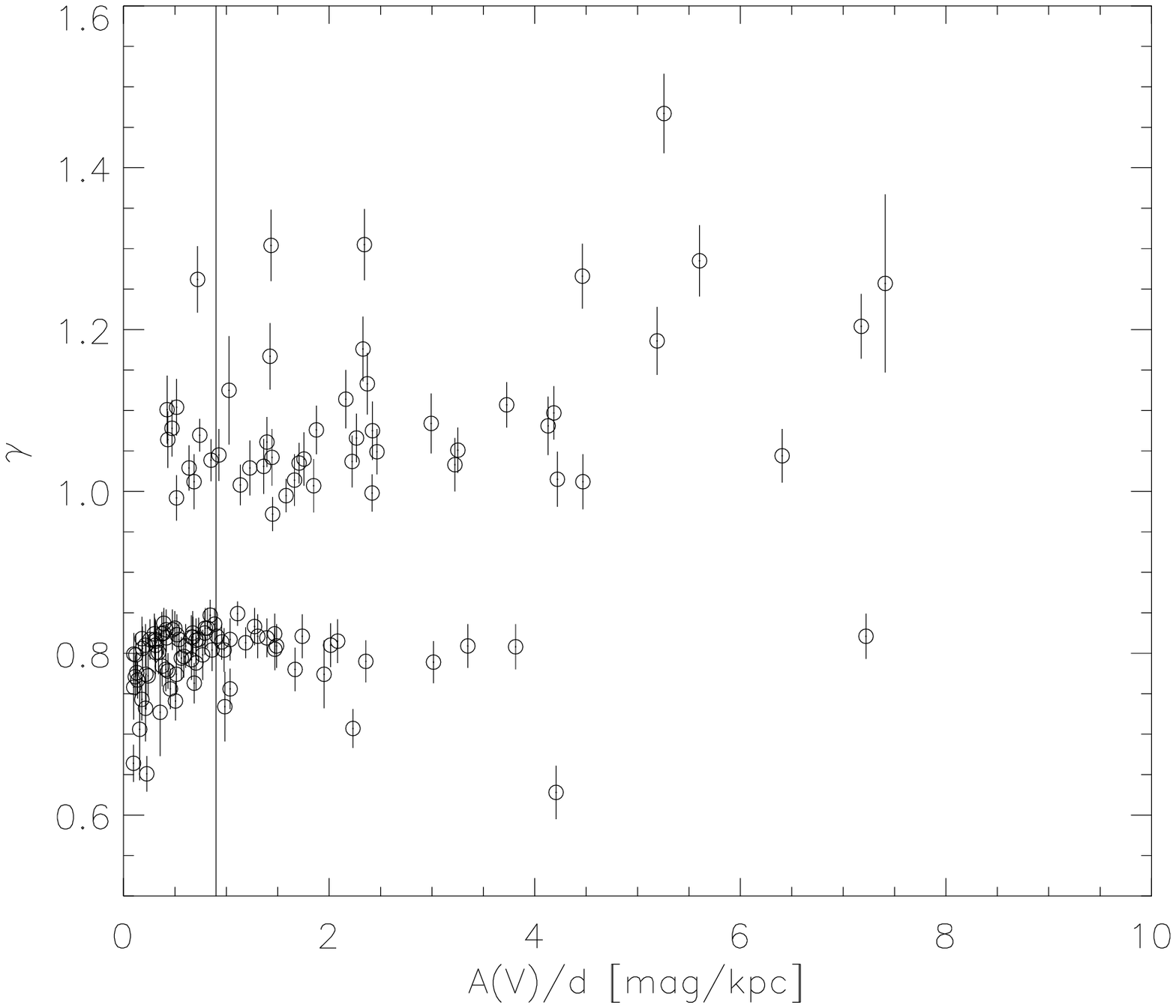}
\end{center}
\caption{Bump width, $\gamma$, vs. density.  The solid line is the cutoff
between dense and diffuse lines of sight.  Left panel: all sightlines
in database.  Right panel: Sightlines beyond 3$\sigma$ of the mean.
In order to make the correlation easier 
to see, those lines of sight that agreed with the mean were 
removed; the remainder are shown in the right hand panel.  
In both, the line at $A_{V}/d$ = 0.9 
indicates the cutoff between dense and diffuse lines of sight. 
\label{fig:gamma_avonr}}
\end{figure}

The width of the bump shows real variation and environmental
dependence.  Values for $\gamma$ ranged from 0.63$\pm$0.03 $\mu$m$^{-1}$~(HD 24263) 
to 1.47$\pm$0.05 $\mu$m$^{-1}$~(HD 29647).  This is a wider range than that 
reported by FM86 ($\gamma$ =0.77$\pm$0.09 - 1.25$\pm$0.07 
$\mu$m$^{-1}$), perhaps reflecting the larger volume of the Galaxy 
and wider range of environments covered by this study. 
The average value in our sample is $\gamma=$0.92$\pm$0.12 $\mu$m$^{-1}$.  
HD 29647 and HD 62542 have the broadest bumps, with
$\gamma$=1.467$\pm$0.049 $\mu$m$^{-1}$ and $\gamma=1.304\pm0.04$ $\mu$m$^{-1}$, 
respectively.  The average $\gamma$ for dense and diffuse sightlines as defined 
above do not differ significantly, as $\gamma_{dense}^{avg}$ = 0.95
$\pm$0.04 $\mu$m$^{-1}$~and $\gamma_{diffuse}^{avg}$ = 0.87 $\pm$0.03 
$\mu$m$^{-1}$.  This likely reflects the density parameter's inability 
to sort out high versus low densities over long distances, as discussed
previously.  A z-test (i.e., Naiman, Rosenfeld, \& Zirkel 1983) was done 
on the diffuse and dense $\gamma$ subsets.  A z-test is essentially the 
same as a t-test, as it is used to determine if two different datasets have 
significantly different mean values, but is used for large samples.  As with 
the t-test, the resulting significance of 
such a test ranges from 0 (the samples have significantly different 
means) to 1 (the samples have essentially the same means).  For example,
a significance of a few hundredths indicates that two samples have 
significantly different means.  For the dense and diffuse sightlines, the 
resulting significance was $\sim$ 10$^{-6}$, which implies that there is 
a very significant difference between the mean $\gamma$ values of the dense and 
diffuse subsets.  Fig. \ref{fig:gamma_avonr} shows a plot of all of the 
sightlines and their environmental dependence, and a plot of sightlines which 
are more than 3$\sigma$ from the mean value of $\gamma$.
It can be seen that there is a clear trend that the narrowest bumps tend
to be found along sightlines with $A_{V}/d<$0.9 mag/kpc, the ``diffuse'' 
subset.  These results strengthen those of FM86 and Cardelli
\& Clayton (1991), who found that lines of sight that
passed through bright nebulosities (``diffuse'' sightlines) had 
narrower bumps than those that passed through dark clouds. 

\begin{figure}[th]
\begin{center}
\includegraphics[scale=.4]{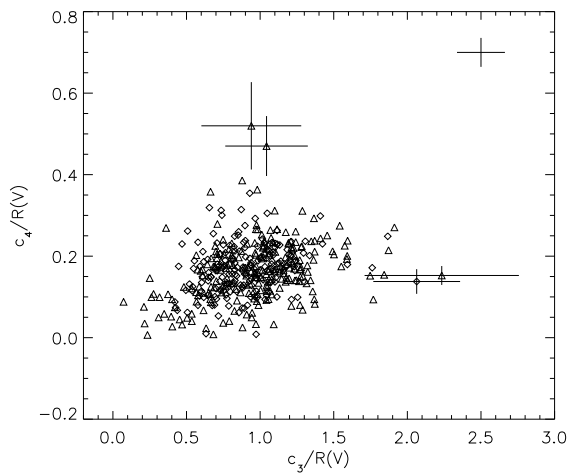}
\end{center}
\caption{FUV curvature, c$_{4}$/R$_V$, vs. bump height, c$_{3}$/R$_V$.  
Symbols are the same as defined in Fig. 2.
Representative error bars are indicated. 
The bump height (c$_{3}$/R$_V$) and
FUV curvature (c$_{4}$/R$_V$) are plotted with respect to bump
width. Values of $\gamma$
were split into three categories, each roughly pertaining to
environment, with diamonds indicating $\gamma < 0.9$, triangles
for $ 0.9< \gamma < 1.1$, and squares for $\gamma > 1.1$.
\label{fig:c4_c3}}
\end{figure}

Fig. \ref{fig:c4_c3} compares c$_{4}$/R(V) and c$_{3}$/R(V), 
with respect to $\gamma$.  It can be seen that as the 
bump height increases, the range of values of the FUV rise 
increases, and the sightlines with the widest bumps tend to 
have higher values of c$_3$/R(V), all of which has been reported 
previously by FM88.  This is likely a reflection of the 
dependence of c$_{3}$/R(V) and $\gamma$ on R$_{V}^{-1}$.  
No similar trends can be seen between c$_{4}$/R(V) and $\gamma$, 
as was discussed earlier. 

\begin{figure}[th]
\begin{center}
\includegraphics[scale=.4]{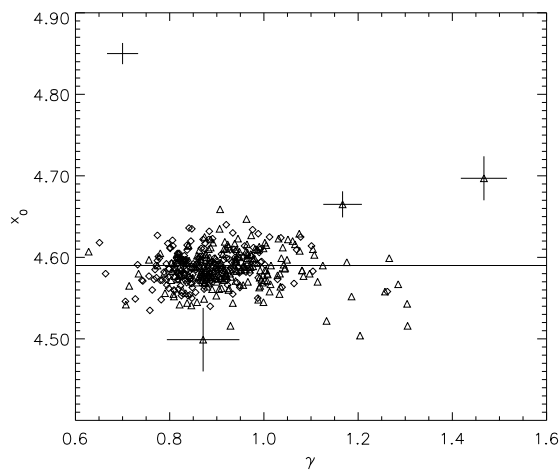}
\end{center}
\caption{Bump central wavelength, x$_{0}$, vs. bump width, $\gamma$.
The solid line is the average.  
Symbols are the same as defined in Fig. 2.
Representative error bars are
indicated. \label{fig:x0_gamma}}
\end{figure}

Fig. \ref{fig:x0_gamma} plots $\gamma$ vs. x$_{0}$, the bump 
central wavelength, for the sample.  This figure shows 
that while $\gamma$ varies significantly from one sightline to 
another, x$_{0}$ does not.  The 
extremely narrow range of x$_{0}$ agrees with FM86,
who found that the mean x$_{0}$ = 4.60 $\mu$m$^{-1}$.
They find
extreme values of x$_{0}$ are within $\pm$0.04 $\mu$m$^{-1}$ of the mean.
Neither HD 62542 nor HD 29647 were included in the FM86 sample.
JG93 found a similar value of $<$x$_{0}> = 4.58 \pm 0.01$ $\mu$m$^{-1}$ for the
Aiello et al. (1988) sample.  For the much larger sample studied here, 
$<$x$_{0}>$ = 4.59$\pm$0.01 $\mu$m$^{-1}$, with values ranging from 
4.50$\pm$0.04 $\mu$m$^{-1}$~ (HD 145792) to 4.70$\pm$0.03 $\mu$m$^{-1}$~(HD 29647).

With the possible exception of HD 29647, it appears that x$_{0}$
does not vary from sightline to sightline.  Cardelli \& Savage (1988) found 
that the 2175 \AA\ bumps toward both HD 62542 and HD 29647 were significantly 
shifted to shorter wavelengths. 
However, we find x$_{0} = 4.54 \pm$ 0.03 $\mu$m$^{-1}$ for HD 62542.
Cardelli \& Savage (1988) found x$_{0}$= 4.74$\pm$0.03 $\mu$m$^{-1}$, and in
the extinction curve shown in their work, the bump is visibly
shifted blueward.  In an effort to reproduce their results, the same
IUE spectra and UV comparison star 
were used to construct an extinction curve.  This curve
was then fit over three ranges; the original FM90 range (3.3 -
8.7 $\mu$m$^{-1}$), from 3.7 - 8.0 $\mu$m$^{-1}$, and from 3.7 - 8.7 $\mu$m$^{-1}$
excluding the region around Ly$\alpha$.  None of these ranges
produced a shifted bump. While the spectra used by Cardelli \& 
Savage (1988) are the same as used here, their spectra were reduced 
and calibrated
with a different software package than those in the final 
archive.  

In our sample of 417 sightlines, only one was shifted as 
much as 3 $\sigma$ beyond the mean. We find x$_{0}$ = 
4.70$\pm$0.03 $\mu$m$^{-1}$ for HD 29647, the same value 
found by Cardelli \& Savage (1988). The other two
fitting ranges that were considered when testing for HD 62542's 
bump shift were applied to HD 29647.  
The values of x$_{0}$ that were found were within
1 $\sigma$ of the mean. Thus, it is quite possible
that the shift in the HD 29647 bump also is fit dependent.
Another sightline, lying close to that of HD 29647 has been studied  
(Clayton et al. 2003).  Both sightlines pass through the Taurus 
Dark Cloud.  The second sightline, toward HD 283809, shows no 
evidence of a shifted bump. 

If the carrier is
small graphite grains, then $\gamma \propto 1/a$, where
$a$ is the grain radius, while
x$_{0}$ is not a function of size (Bohren \& Huffman 1983; Hecht 1986).
Hecht concluded that some fraction of small
($a < 50 $\AA) carbonaceous grains are bare and 
these are responsible for the bump.
In this scenario, the bump width is dependent on the
temperature and size distribution of these grains.  The
remaining small carbon grains are hydrogenated, which
would suppress the bump (Hecht 1986).  One 
drawback to this model was
that it required an overabundance of small grains in dense regions
relative to diffuse regions (FM86; Sorrell 1990), since a large
value of $\gamma$ implies small grain size in this scenario.
This conflicts with the observation that dense regions tend to have higher
values of R$_V$, indicating the prevalence of large grains.
However, since x$_{0}$
is not dependent on R$_V$ (or anything else), the grains which produce
the bump may form a separate population from those which are responsible
for variations in R$_V$ (CCM).  On the other hand, it can
be seen in Fig. \ref{fig:CCM_karl} that there is a weak
correlation between $\gamma$ and R$_V$$^{-1}$.  This agrees
with the finding that $\gamma$ is environment dependent.
In 1990, Sorrell expanded upon Hecht's (1986) work, suggesting that
the accretion of hydrogen on larger
graphite grains with 60 \AA$ < a < 80$ \AA~ was responsible
for the variations in $\gamma$.
He suggested that hydrogenation of  grains in dense regions
broadens the bump, without affecting x$_{0}$.

Mathis (1994) considered diamond, amorphous carbon (AMC), water ice,
and PAH mantles on graphite cores.  He found that the
diamond coating tended to shift x$_{0}$
to longer much wavelengths.  A thin coating
of amorphous carbon  yields
x$_{0}$ =4.57 $\mu$m$^{-1}$ but produces a low $\gamma$= 0.84 $\mu$m$^{-1}$.  
Increasing the thickness of mantle does not increase $\gamma$
to values that are observed; a thicker mantle only 
broadens the bump to $\gamma$=0.93 $\mu$m$^{-1}$.  Hydrogenated
amorphous carbon mantles are even less suitable,
as they shift x$_{0}$ to longer wavelengths
while having even less of an impact on $\gamma$ than AMC did.
Water ice mantles may broaden $\gamma$, but only one
sightline in our sample (HD 29647) has the
telltale 3.07 $\mu$m ice feature (Goebel 1983).  
Warren (1984) showed that ice mantles do not
affect the bump because there is almost no
absorption at wavelengths near 2175 \AA.
Neutral PAHs are also possible bump grain candidates, as they can
produce both x$_{0}$ and $\gamma$ consistent
with observations, with x$_{0} \approx$ 4.61 $\mu$m$^{-1}$~
for $\gamma$= 1.0 $\mu$m$^{-1}$~(Mathis 1994).  However, these values were found by
assuming that the optical constants of isolated
PAHs are similar to those comprising the mantle
and by ignoring impurities which might alter the
PAH optical properties.  Another drawback is that
neutral PAHs should have absorption features in the UV
which are not seen (Clayton et al. 2003), though 
Joblin et al. (1992)
showed that this could be masked by averaging over a
distribution of PAHs of different sizes.  Also,
it is unlikely that many PAHs are neutral in the 
diffuse ISM (e.g., LePage et al. 2003). 

\subsection{Deviations from CCM}

\begin{figure}[th]
\begin{center}
\includegraphics[scale=.5]{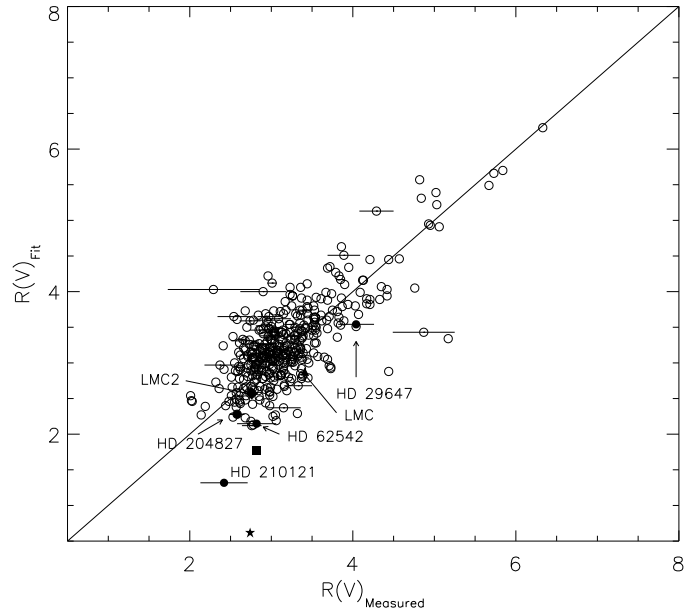}
\end{center}
\caption{The best-fit R$_V$ values, found through $\chi^{2}$ minimization,
compared to R$_V$ values found through IR photometry.  A line of slope
unity has been overlayed.  Four lines of sight with unusual extinction
are indicated (filled circles).  Lines of sight where the best-fit
R$_V$ did not agree with the measured value are indicated with error bars.
The average LMC (filled triangle), LMC2 Supershell (filled diamond), and 
SMC (filled star) are plotted.  From Clayton et al. (2000), the average 
low-density sightline for a Galactic region with extinction similar 
to that of the Magellanic Clouds (the ``SD region'') is also shown (filled 
square).  
\label{fig:rv_test}}
\end{figure}

Adherence to the R$_V$-dependent extinction law of CCM in our sample was also
examined. 
In order to do this, we found the CCM curve which best fit each 
extinction curve. 
The extinction curves were fit with a CCM
curve using a standard IDL curve fitting routine that minimized 
the $\chi^{2}$.  The values of R$_V$ found this way were compared to
those calculated using IR photometry as is shown in
Fig.\ \ref{fig:rv_test}.  While there is some scatter, there is 
generally good agreement
between the two values, with the fitted R$_V$ being within
3$\sigma$ of the calculated value for 93\% of the sample.
This indicates that the vast majority of sightlines agree with CCM
within the measurement uncertainties.
The near absence of non-CCM sightlines in the Galaxy
could be attributed to the IUE dataset
somehow favoring lines of sight through CCM-type dust environments. 
IUE was a small telescope with limited dynamic range. As a result,
there is little information
on grain parameters deep inside dark clouds.  
The dust columns sampled in this study all have A$_V <$ 5 mag. 

Next we measured deviations of each measured extinction curve 
from its best fit CCM curve. 
We followed the method of
Mathis \& Cardelli (1992; hereafter MC92), in which the authors specified the
deviation at various wavelengths to be:

\begin{displaymath}
  \delta(x_{i}) = [A_{\lambda}/A_{V}]_{i} - [A_{\lambda}/A_{V}]_{CCM}
\end{displaymath}

\begin{figure}[th]
\begin{center}
\includegraphics[scale=.35]{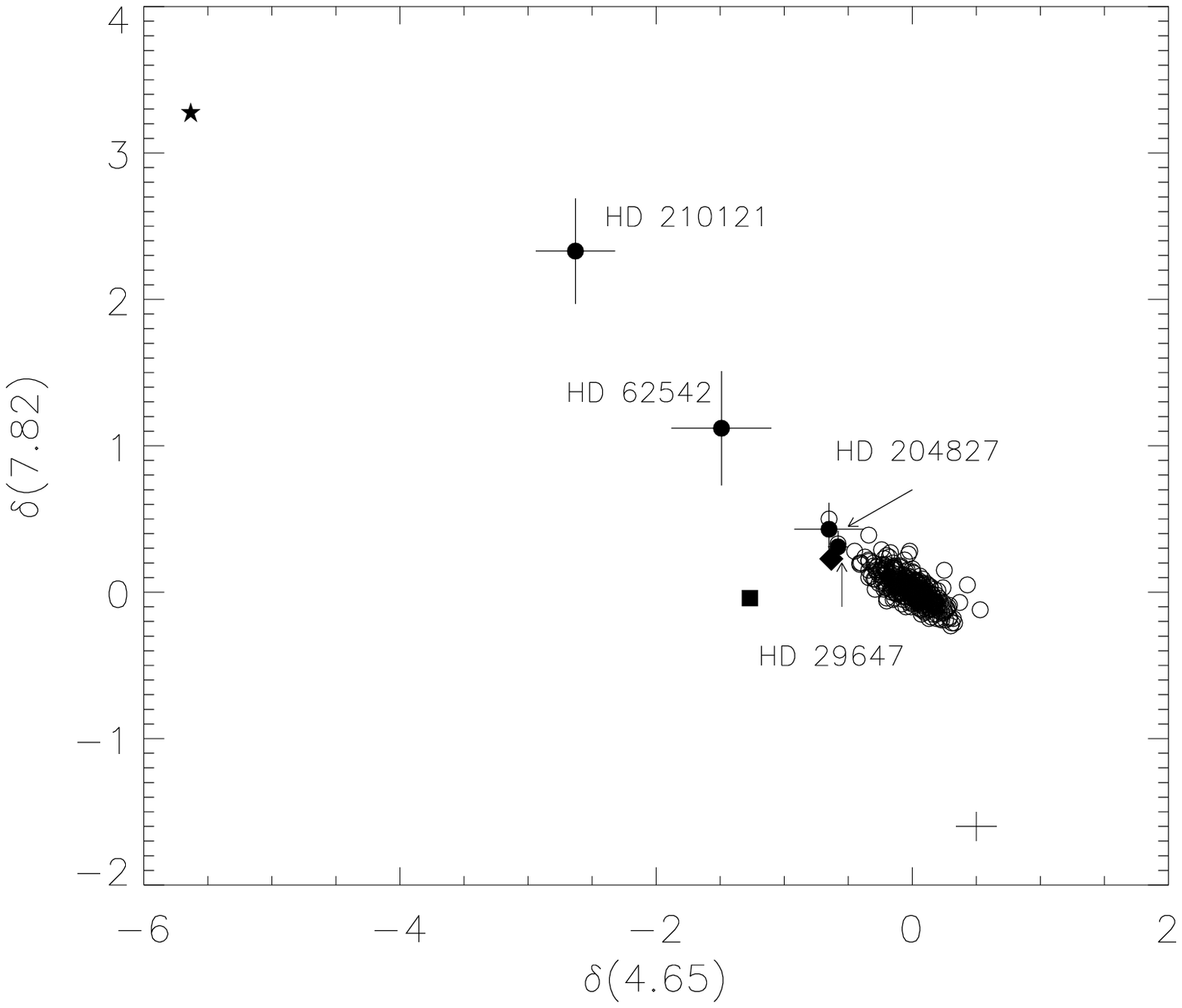}
\includegraphics[scale=.35]{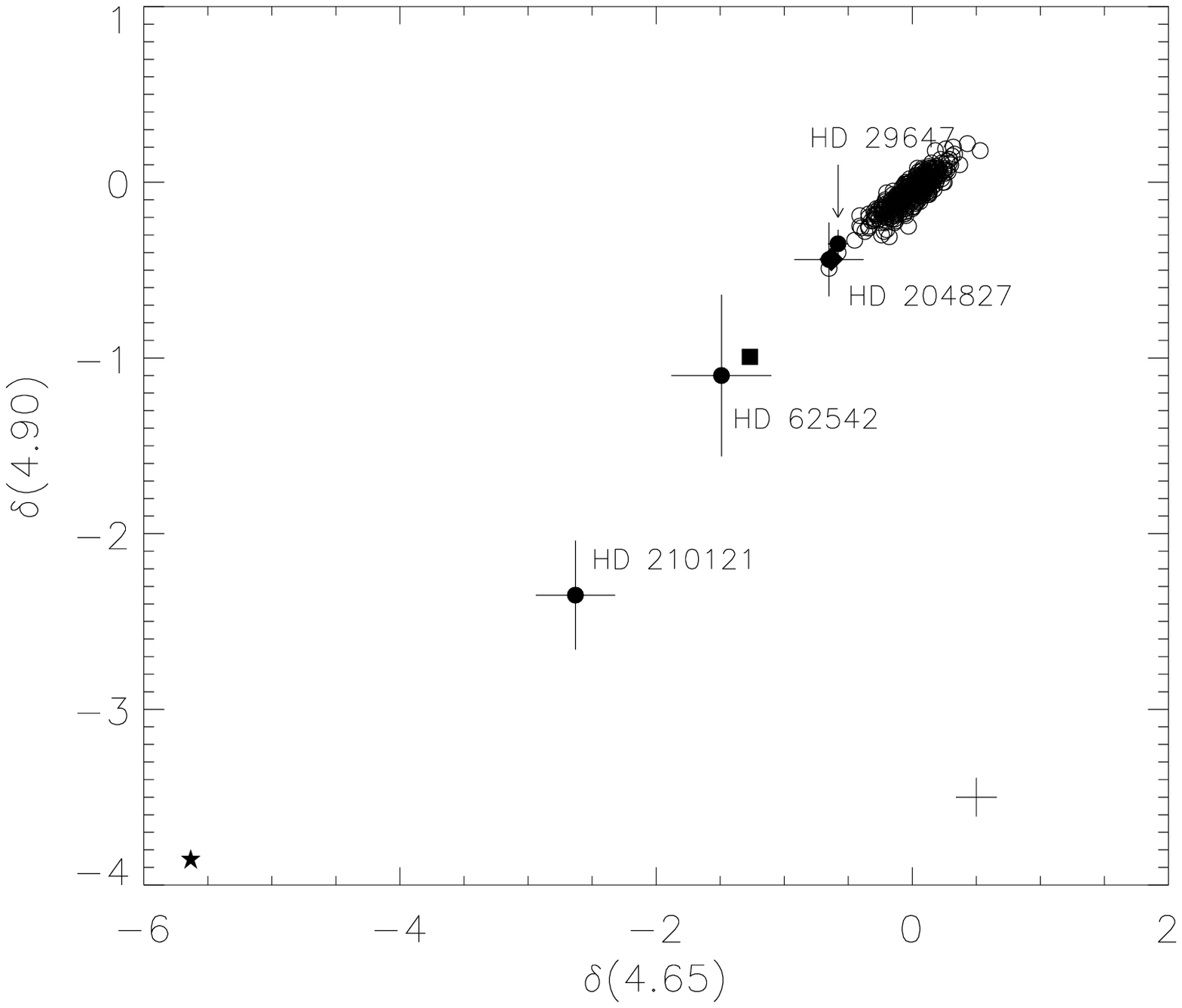}
\includegraphics[scale=.35]{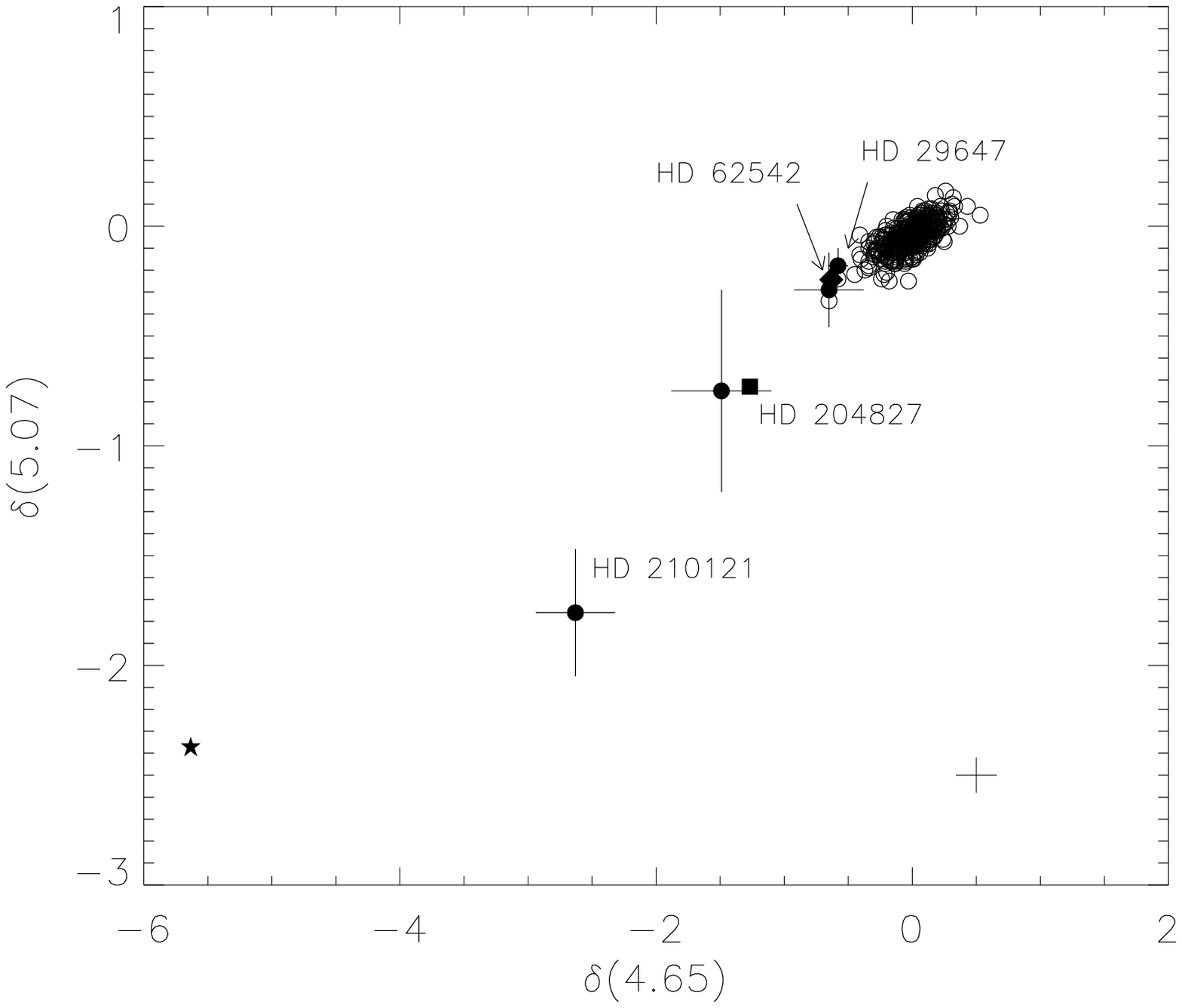}
\includegraphics[scale=.35]{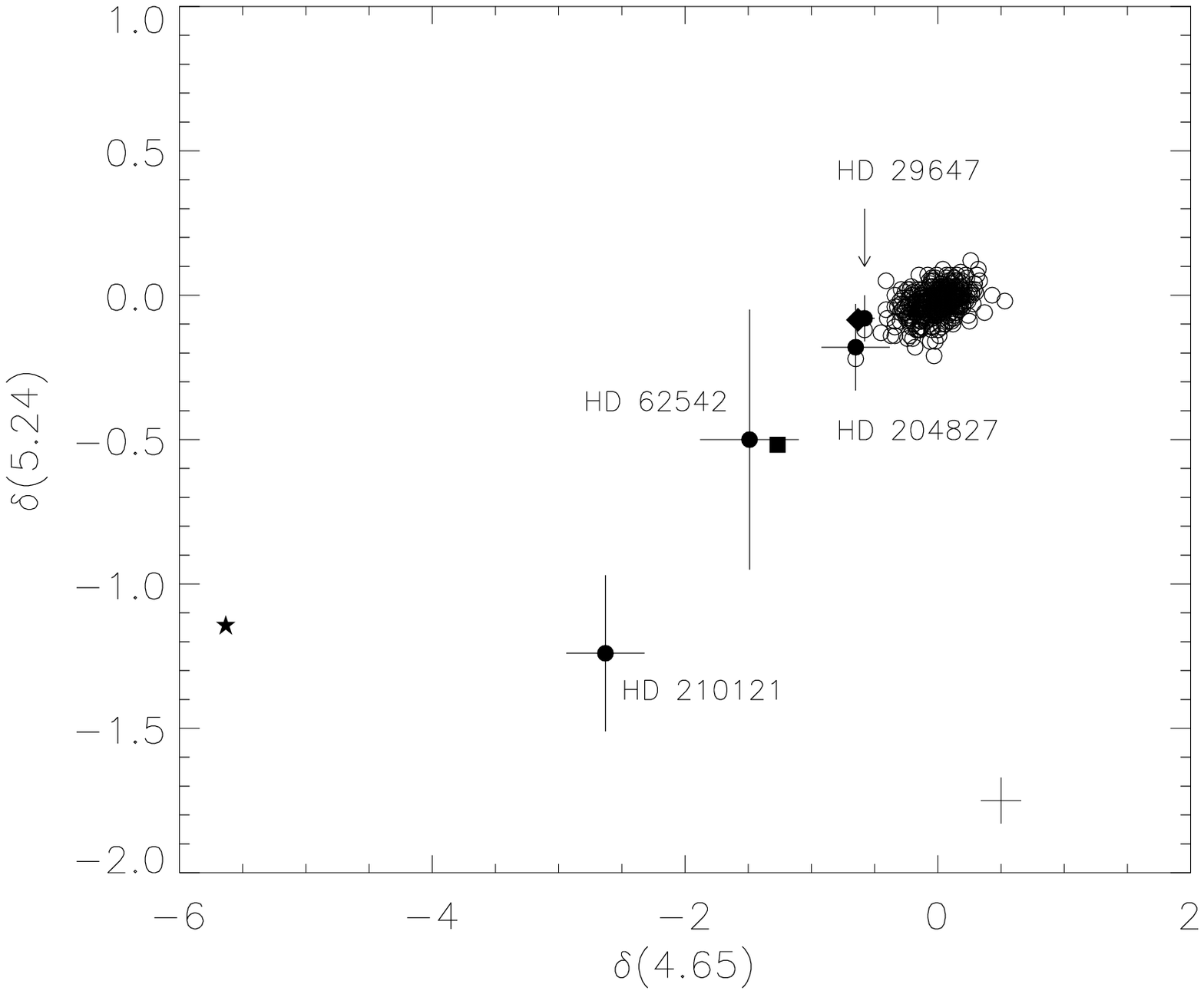}
\includegraphics[scale=.35]{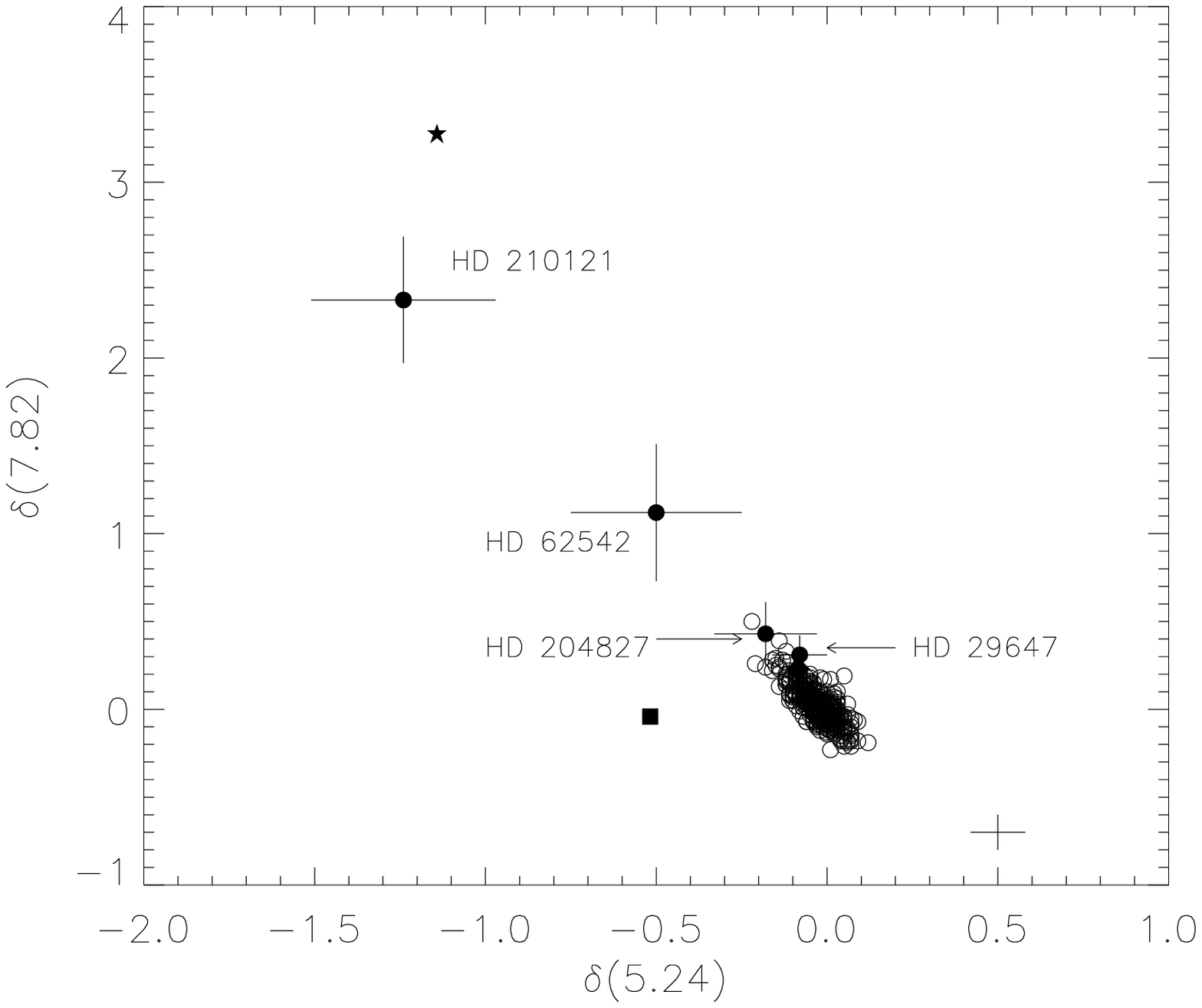}
\end{center}
\caption{The deviations from R$_V$ law at various wavelengths.
Representative error bars are indicated.  Four Galactic lines of 
sight with unusual extinction are named.  The SD region average sightline 
(Clayton et al. 2000; filled square), average LMC (filled triangle), 
LMC2 Supershell (filled diamond), and SMC (filled star) are also 
plotted. 
\label{fig:deviations}}
\end{figure}

\noindent
Plots are shown of these deviations in Fig. \ref{fig:deviations}
at 4.65, 4.90, 5.07, 5.24,
and 7.82 $\mu$m$^{-1}$.
The calculated and best-fit CCM extinction curves agreed 
within 2$\sigma$ of
each other for all but four of the sightlines, 
HD 210121, HD 204827, HD 62542, and
HD 29647, which are deviant at several wavelengths. Their 
deviations are shown in Table \ref{tab:deviations_list}. 

HD 204827 is an intriguing sightline. It has been shown
that when a foreground reddening component has been removed, 
its extinction curve resembles that of the SMC, even
though the environment of the dust in the HD 204827 cloud is quite
different from that seen in the SMC sightlines (Valencic et al. 2003). 
A similar curve has been found toward HD 283809 (Whittet et al. 2004).
The dust local to HD 204827
has high density. The HD 204827 dust cloud resembles a molecular 
cloud more than the diffuse ISM.  The HD 204827 cloud is very rich 
in carbon molecules, showing large column densities of C$_2$, C$_3$, 
CH, and CN
(Oka et al. 2003; Thorburn et al. 2003).  
The sightline to HD 204827 samples dust swept up by a 
supernova or hot star winds (Patel 1998). 
In these respects, the HD 204827
dust is quite similar to the sightline toward HD 62542 (Cardelli
\& Savage 1988). Its dust is also
rich in carbon molecules. The sightline to HD 62542 also
lies on the edge of material swept up by a stellar wind bubble.
The other two non-CCM, weak bump, steep far-UV sightlines in the Galaxy,
HD 29647, and HD 210121, are also associated with dense
clouds (Cardelli \& Savage 1988; Larson, Whittet, \& Hough 1996;
Cardelli \& Wallerstein 1989).  The dust in the
molecular cloud associated with HD 210121 is likely to have been
processed as it was propelled into the halo during a Galactic fountain
or other event.  The final sightline, toward HD 29647, is sampling 
dust in a quiescent dense cloud.
The steep far-UV extinction in these clouds helps shield the molecules
in these clouds from dissociating UV radiation leading to larger
column densities (Mathis 1990).  

\begin{figure}[th]
\begin{center}
\includegraphics[scale=.5]{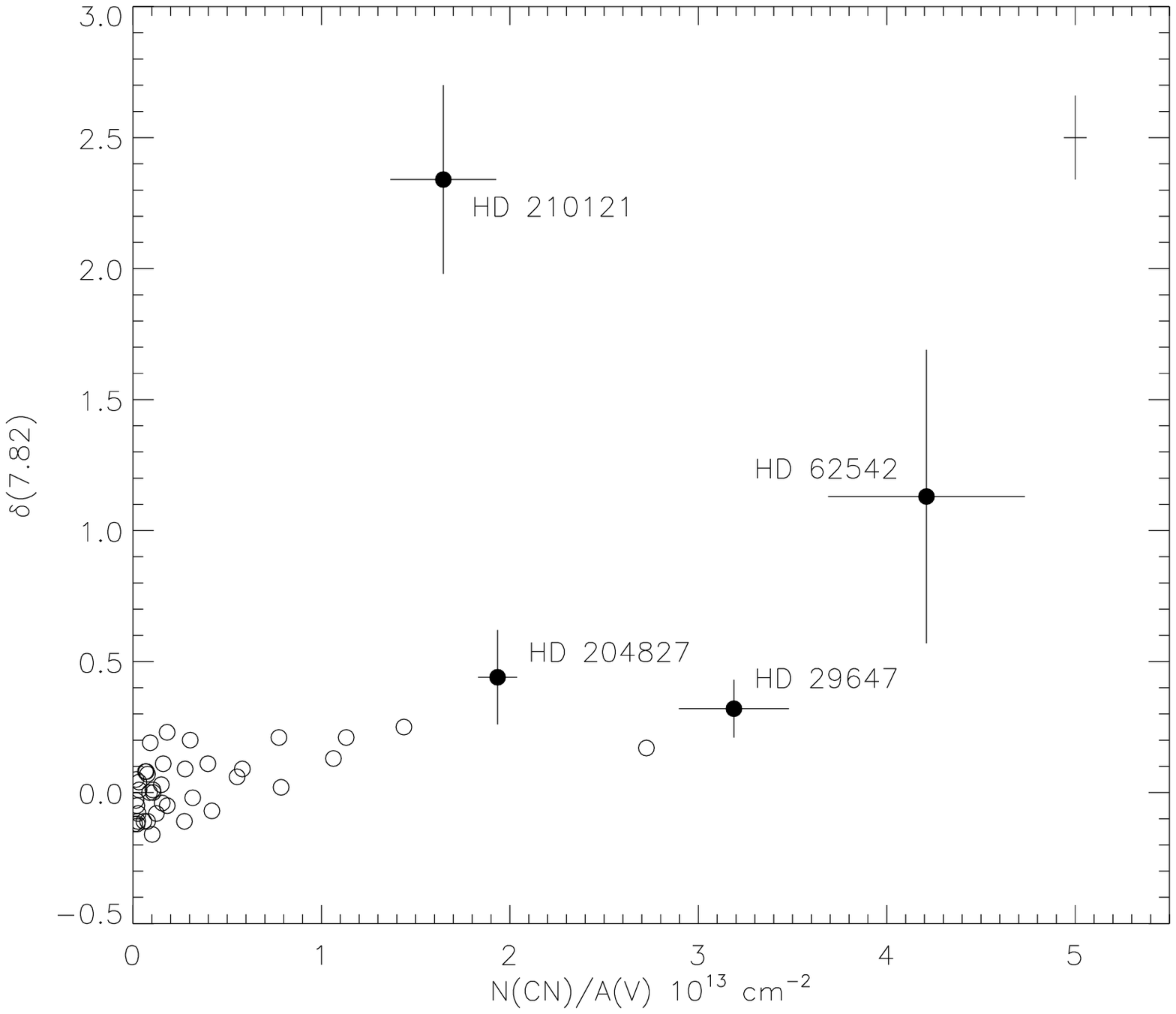}
\includegraphics[scale=.5]{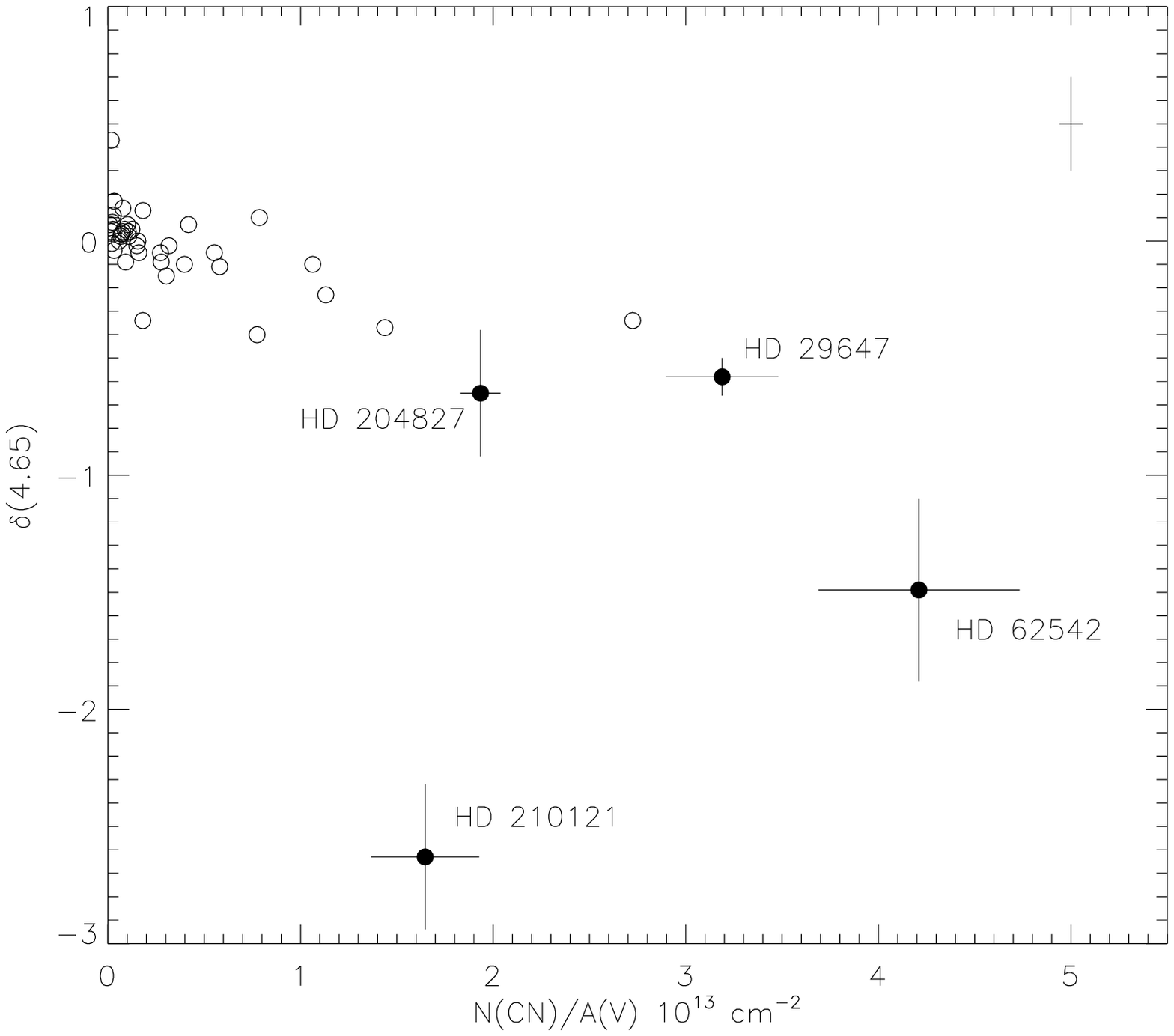}
\end{center}
\caption{Deviation at 7.82 $\mu$m$^{-1}$ and 4.65 $\mu$m$^{-1}$ versus 
abundance of CN.  Values of N(CN) are from Federman (1994) and Oka et 
al. (2003).  Unusual Galactic sightlines are indicated with solid 
cirles.  
\label{fig:delta782_NCN_sig2}}
\end{figure}

Fig. \ref{fig:delta782_NCN_sig2} shows the deviations, $\delta(7.82)$ 
and $\delta(4.65)$, 
plotted against the column density of CN along the sightline divided
by A$_V$ for 46 sightlines from our sample, including the four non-CCM 
sightlines discussed above (Federman 1994; Oka et al. 2003).  
CN can be used as an
indicator of relatively dense regions in diffuse, molecule-rich clouds
(Joseph et al. 1986; Gredel et al. 2002).  The average $\delta(7.82)$
and N(CN)/A$_{V}$ (10$^{13}$cm$^{-2}$ mag$^{-1}$) for the 
42 CCM-like sightlines were
0.03$\pm$0.03 and 0.33$\pm$0.01, respectively. 
For the four non-CCM sightlines,
these values were 1.06$\pm$0.15 and 2.75$\pm$0.15.   
These four sightlines have both high N(CN)/A$_{V}$ and sigificant
deviation from CCM at $x$=7.82 compared to the CCM-like sample. 
Similarly, Burgh et al. (2000) showed that sightlines with steep 
FUV rises tend to have high N(CO)/E(B-V), though they did not 
notice any correlating weakness in the bump strength.  

Virtually all
of the non-CCM sightlines known today are in the (diffuse)
Magellanic Clouds, not in dense Galactic clouds (e.g., Gordon et al. 2003).
Clayton et al. (2000), found that the average of seven 
low density, low reddening sightlines showed 
an extinction curve very similar
to the LMC.  These 7 stars were found to be behind
gas that showed signs of being subjected to shocks  
The deviations from CCM for the average of
these 7 sightlines as well as for the SMC and LMC average curves were
also found and are included in Figs. \ref{fig:deviations} 
and \ref{fig:rv_test}.   
Similar deviations from CCM may arise in various dust environments.

\subsection{The Updated CCM Law} 

The new extinction curves and values of R$_{V}$ allow us to revise 
the CCM law.  This was done by following the method described in 
CCM; that is, by plotting A$_{\lambda}$/A$_{V}$ against R$_{V}^{-1}$ 
for all wavelengths, then performing a linear least-squares fit of 
the resulting correlation.  Thus, for all wavelengths, the extinction 
A$_{\lambda}$/A$_{V} = a(x) + b(x)/$R$_V$.  Then, $a(x)$ and $b(x)$ 
were plotted against $x$ and fit with a polynomial expression by 
minimizing $\chi^{2}$.  It was found that for 3.3 $\ge x \ge 8.0 ~\mu$m$^{-1}$,

\begin{eqnarray*}
a(x) = 1.808 - 0.215x - \frac{0.134}{(x-4.558)^2 + 0.566} + F_{a}(x)  \\
b(x) = -2.350 + 1.403x + \frac{1.103}{(x-4.587)^2 + 0.263} + F_{b}(x)
\end{eqnarray*} 
   
where, for $x < 5.9 \mu$m$^{-1}$, 

\begin{displaymath} 
  F_{a}(x) = F_{b}(x) = 0.0
\end{displaymath}

and for $5.9 \le x \le 8.0 \mu$m$^{-1}$,

\begin{eqnarray*} 
  F_{a}(x) = -0.0077(x-5.9)^2 - 0.0003(x-5.9)^3  \\
  F_{b}(x) = 0.2060(x-5.9)^2 - 0.0550(x-5.9)^3.  
\end{eqnarray*} 

\begin{figure}[tb]
\begin{center}
\includegraphics[scale=.5]{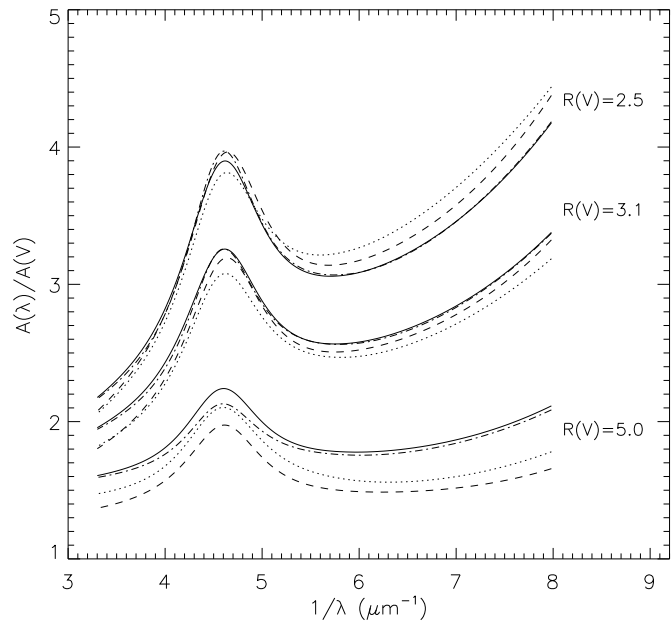}
\end{center}
\caption{A comparison of extinction laws.  Solid line: the extinction law 
derived in this work; dashed line: the original CCM law; dash-dot line: 
the curve constructed with the FM parameter averages found in this work; dotted 
line:  the curve constructed with the FM parameters suggested by Fitzpatrick 1999.   
\label{fig:lots_of_lines}}
\end{figure}

A comparison of extinction laws are plotted in Fig. 
\ref{fig:lots_of_lines}.  The most noticeable difference between 
the law found here and the original CCM law, the increase in the 
zero-point of the UV extinction, is a consequence of the slightly 
higher value of the first term in $a(x)$.  In 1988, CCM also found
a higher value for this term (1.802) than in 1989 (1.752), when 
they found that the lower value, which was within the dispersion 
in the data, was needed for the curve to smoothly join the optical 
extinction data.  Fig. \ref{fig:lots_of_lines} also shows the resulting 
curve when the average FM parameter values are used to construct 
extinction curves at different values of R$_{V}$, in a fashion similar 
to Fitzpatrick (1999).  The two methods 
of curve construction yield curves that are within 5\% of each other 
for 2.5 $\le$ R$_{V} \le$ 5.0 over the entire wavelength range.  For all 
values of R$_{V}$, the difference is primarily in the bump height, or 
the FM parameter c$_{3}$.  This is most noticeable for 5.0 $<$ R$_{V} 
\le$ 6.0; here, the curves agree to within 7\%, with the curve found 
through the average FM parameters having a weaker bump than predicted by 
the either the revised or original CCM law.  This can also be seen to 
a certain extent in Fig. \ref{fig:CCM_karl}.  A weakened bump at high 
R$_{V}$ may be expected from Whittet et al.'s (2004) work, which 
suggests that the bump carrier may be destroyed in dark clouds. 
Both the revised CCM curves and the 
average FM curves are within about 5\% of the original CCM law for 
2.5 $\le$ R$_{V} \le$ 3.5.  However, for increasing values of R$_{V}$, 
the difference between the curves increases as well, reaching 20\% at 
R$_{V} \sim 5.0$.  That the original CCM law agreed so well with the 
extinction curves examined here is an indicator of the sizes of the 
uncertainties on the data.  Changes in the value of the extinction 
zero-point is an important factor.  If the first term in $a(x)$ is 
decreased by 0.20, the new CCM law and the original curve agree to 
within 15\% for 2.5 $\le$ R$_{V}$ $\le$ 6.0 for $x < 7.0 ~\mu$m$^{-1}$, 
and within about 20\% for $x \ge 7.4 ~\mu$m$^{-1}$. 
The dotted curve, constructed from Fitzpatrick's (1999) suggested values 
of FM parameters with only c$_2$ (and through it, c$_1$) dependent on 
R$_{V}$, illustrate further the dependences of some FM parameters 
on R$_V$.

\section{Conclusions}

We have constructed a homogeneous database
of UV extinction curves, all of which have been fit by
the FM relation and for which values of R$_V$ have been
calculated.  All extinction
curves and FM parameters have been normalized to A$_{V}$,
rather than E(B-V), so that relationships between
parameters may be more easily seen.  This is the largest 
and most comprehensive database of extinction curves yet 
constructed, sampling 
a wide variety of environments.  It contains much longer 
sightlines than those included in previous studies (30 
sightlines have d $>$ 5 kpc), thus greatly increasing the 
volume of the Galaxy sampled compared to previous works. Over 150
sightlines have d $>$ 2 kpc, so the regions of the Galaxy
beyond the solar neighborhood are well represented.

The main results of this study are:

\vspace{0.5cm}

\noindent
(1.) The CCM extinction relation accurately describes
the diffuse Galactic ISM in virtually all cases.  Out of 
417 sightlines, only 4 deviated consistently from their 
CCM extinction curves by more than 3$\sigma$.  
This indicates that the physical processes which give 
rise to grain populations that have CCM-like extinction 
dominate the ISM, and thus, the quantity R$_V$ can accurately 
describe the UV extinction for most sightlines.  This implies 
that the grain populations responsible for different components
of the extinction curve are being processed efficiently
and systematically along most sightlines. 

The new curves and R$_{V}$ values allow for an updated CCM law 
to be made.  The original and updated versions are within 5\% of 
each other for 2.5 $\le$ R$_{V} \le 3.5$ over the full wavelength 
range covered, 3.3 $\le$ x $\le 8.0 ~\mu$m$^{-1}$.  At higher 
values of R$_{V}$, the differences in the extinction zero-point reduce
the agreement in the curves to about 20\%. 

\vspace{0.5cm}

\noindent
(2.) The bump width has a strong environmental
dependence, with narrow bumps favoring diffuse
sightlines, and broad bumps favoring dense sightlines.
Very broad bumps are rare, as only 21 out of
417 sightlines had $\gamma >$ 1.1.  

\vspace{0.5cm}

\noindent
(3.) The central wavelength of the bump is invariant
and may be regarded as a constant,
with x$_{0}$=4.59$\pm$0.01.  Unlike the other parts of
the UV extinction curve, this parameter does not respond to different
environments. The invariance of x$_{0}$, along with the observed
variations in $\gamma$, put strong constraints on possible bump
grain and mantle materials.

\vspace{0.5cm}

\noindent
(4.)  While there is evidence for shock processing in three of
the four non-CCM sightlines, the common denominator is that all
four sightlines have dense, molecule-rich clouds.  They also had 
weak bumps and strong FUV extinction for their measured R$_V$ CCM 
curves, especially those which pass through cold, quiescent regions.
The weakened bump may reflect processing which
modifies or destroys the bump carrier in dark clouds (Whittet
et al. 2004). This is seen in the bumpless dust toward HD 204827
and HD 283809.  Other dense sightlines with strong FUV extinction,
HD 204827, HD 62542, and HD 210121, pass through
dense clouds that may have been exposed to shocks or
strong UV radiation that disrupt large grains,
possibly resulting in a size distribution that is
skewed toward small grains.

Together, the sightlines in the Galaxy and the Magellanic Clouds 
suggest that similar extinction properties may arise from very 
different environments.

\vspace{0.5cm}
Financial support was provided by the Louisiana Board of Regents,
BoRSF, under agreement NASA/LEQSF(1996-2001)-LaSPACE-01 or
NASA/LEQSF (2001-2005)-LaSPACE and NASA/LaSPACE under grant
NGT5-40115.

\newpage

\begin{table}[tb]
\caption{Comparison of Previous UV Extinction Studies to the Present Work \label{tab:previous_work}}
\begin{center}
\begin{tabular}{cccccc}
\hline \hline
(1) & (2) & (3) &(4)& (5) &(6) \\
Authorship & No. of Sightlines & Instrument & FM & Spectral & R$_{V}$ \\ 
  &  &  &  & Classification & \\
\hline 
Savage et al. 1985 & 1415 & ANS & no & MK & no \\
Aiello et al. 1988 & 115 & IUE & no & UV & no \\
FM90 & 78 & IUE & yes & UV & no \\ 
Papaj et al. 1991 & 166 & TD-1 & no & UV & no \\
JG93 & 115 & IUE & yes & UV & no \\ 
Barbaro et al. 2001 & 252 & ANS & yes & MK & yes \\
This Work & 417 & IUE & yes & UV & yes \\
\hline
\end{tabular}
\end{center}
\end{table}

\clearpage

\begin{sidewaystable}[tb]
\caption{Photometry and Sources Used in Present Work \label{tab:phot}}
\begin{center}
\begin{tabular}{lllllllllccc}
\hline \hline
Name & Spectral & UV & U  & B & V  & J &  H &  K & UBV & U & Non-2MASS  \\
 &Type &Type& & & & & & & Source & Source$^\dagger$ & Source \\  
\hline 
HD 14052 & B1Ib & B1Iab & 7.90 & 8.48 & 8.18 & 7.51 & 7.50 & 7.42 & 1 & ... & ... \\ 
         & & & $\pm$0.01 & $\pm$0.01 & $\pm$0.01 & $\pm$0.03 & $\pm$0.03 & $\pm$0.02 &&\\
HD 14357 & B2II & B2Ib & 8.31 & 8.83 & 8.52 & 7.86 & 7.83 & 7.80 & 1 & ... & ... \\
         & & & $\pm$0.01 & $\pm$0.01 & $\pm$0.01 & $\pm$0.03 & $\pm$0.02 & $\pm$0.02 &&\\  
HD 41690 & B1V & B1.5III & 7.31 & 7.93 & 7.71 & 7.20 & 7.23 & 7.20  & 1 &...  & ...\\
         & & & $\pm$0.01 & $\pm$0.01 & $\pm$0.01 & $\pm$0.02 & $\pm$0.02 & $\pm$0.03 &&\\
HD 199216 & B1II &  B2III & 7.05 & 7.51 & 7.03 & 6.05 & 6.03 & 5.97 & 1 &...  & ...\\
          & & & $\pm$0.01 & $\pm$0.01 & $\pm$0.01 & $\pm$0.02 & $\pm$0.03 & $\pm$0.02 &&\\
HD 203938 & B0.5IV & B1.5III & 7.13 & 7.54 & 7.08 & 5.97 & 5.88 & 5.81 &1,2&...  & ...\\
          & & & $\pm$0.01 & $\pm$0.01 & $\pm$0.01 & $\pm$0.02 & $\pm$0.02 & $\pm$0.02 &&\\
HD 235783 & B1Ib & B1Iab & 8.14 & 8.85 & 8.68 & 8.28 & 8.27 & 8.26 & 1 &...  & ...\\
          & & & $\pm$0.01 & $\pm$0.01 & $\pm$0.01 & $\pm$0.03 & $\pm$0.05 & $\pm$0.02 &&\\
\hline
\end{tabular}
\end{center}
$^\dagger$ If not the same as UBV Source.\\
The complete version of this table is in the electronic edition of
the Journal.  The printed edition contains only a sample.
\end{sidewaystable}

\begin{table}[tb]
\caption{ \label{tab:citation_key} Photometry Sources }
\begin{center}
\begin{tabular}{ll}
\hline \hline 
Number & Citation \\
\hline 
1 & Hiltner, W. A. 1956, ApJS, 2, 389 \\ 
2 & Nicolet, B. 1978, A\&AS, 34, 1 \\ 
3 & Schild, R., Garrison, R., \& Hiltner, W. 1983, ApJS, 51, 321 \\ 
4 & Haupt, H., \& Schroll, A. 1974, A\&AS, 15, 311 \\ 
5 & Guetter, H. 1974, PASP, 86, 795 \\ 
\hline
\end{tabular}
\end{center}
The complete version of this table is in the electronic edition of
the Journal.  The printed edition contains only a sample.
\end{table}

\begin{table}[tb]
\caption{General Extinction Characteristics \label{tab:ext_query}}
\begin{center}
\begin{tabular}{lccccccccc}
\hline \hline 
Name & E(B-V) & R$_{V}$ & A$_{V}$ & Distance & Max. Distance & Min. Distance \\
     &        &         &         &  (pc) & (pc) & (pc) \\
\hline 
HD 14052 & 0.49 & 2.99 & 1.47 & 4195.08 & 3322.81 & 6065.88\\ 
         & $\pm$0.04 & $\pm$0.19 & $\pm$0.14 & & & \\ 
HD 14357 & 0.48 & 2.90 & 1.39 & 3817.16 & 2838.24 & 5229.96 \\
         & $\pm$0.04 & $\pm$0.21 & $\pm$0.15 & & & \\
HD 41690 & 0.47 & 2.91 & 1.37 & 820.92 & 602.25 & 1111.31 \\
         & $\pm$0.04 & $\pm$0.22 & $\pm$0.16 & & & \\
HD 199216 & 0.72 & 2.62 & 1.88 & 1567.69 & 1155.49 & 2380.10 \\
          & $\pm$0.04 & $\pm$0.36 & $\pm$0.27 & & & \\
HD 203938 & 0.71 & 3.13 & 2.22 & 496.81 & 362.62 & 672.56 \\
          & $\pm$0.04 & $\pm$0.14 & $\pm$0.16 & & & \\
\hline
\end{tabular}
\end{center}
The complete version of this table is in the electronic edition of
the Journal.  The printed edition contains only a sample.
\end{table}

\begin{table}[tb]
\caption{FM Parameters 
\label{tab:FM_query}}
\begin{center}
\begin{tabular}{lcccccccccccc}
\hline \hline
Name & c$_{1}$/R$_{V}$ + 1.0 & c$_{2}$/R$_{V}$ & c$_{3}$/R$_{V}$ & c$_{4}$/R$_{V}$ & x$_{0}$ & $\gamma$ \\
\hline 
HD 14052 & 1.1856 & 0.2271 & 0.8595 & 0.2124 & 4.5740 & 0.8440 \\ 
         & $\pm$0.3918 & $\pm$0.0258 & $\pm$0.1237 & $\pm$0.0411 & $\pm$0.0140 & $\pm$0.0280 \\ 
HD 14357 & 1.1134 & 0.2745 & 1.0155 & 0.2421 & 4.5830 & 0.8970 \\
         & $\pm$0.1461 & $\pm$0.0282 & $\pm$0.1364 & $\pm$0.0296 & $\pm$0.0050 & $\pm$0.0280 \\
HD 41690 & 1.3412 & 0.2326 & 0.7100 & 0.2790 & 4.6050 & 0.7800 \\
         & $\pm$0.2734 & $\pm$0.0347 & $\pm$0.1193 & $\pm$0.0494 & $\pm$0.0120 & $\pm$0.0270 \\ 
HD 199216 & 0.7992 & 0.2908 & 1.3424 & 0.1408 & 4.5900 & 0.9360 \\
          & $\pm$0.2038 & $\pm$0.0436 & $\pm$0.2244 & $\pm$0.0339 & $\pm$0.0090 & $\pm$0.0340 \\ 
HD 203938 & 1.1581 & 0.1933 & 1.1735 & 0.1371 & 4.5580 & 1.0120 \\
          & $\pm$0.3085 & $\pm$0.0190 & $\pm$0.1534 & $\pm$0.0217 & $\pm$0.0110 & $\pm$0.0340 \\ 
\hline
\end{tabular}
\end{center}
The complete version of this table is in the electronic edition of
the Journal.  The printed edition contains only a sample.
\end{table}

\begin{table}[tb]
\caption{Correlations between FM Parameters \label{tab:params_correl}}
\begin{center}
\begin{tabular}{lcc}
\hline \hline
$r$ & Absolute Deviation & Linear Fit \\
\hline 
0.58 & 0.19 & c$_{3}$ = 2.12 $\gamma$ - 0.95 \\
-0.79 & 0.13 & c$_{1}$ = -2.54 c$_{2}$ + 1.67 \\
0.49 & 0.54 & c$_{3}$ = 1.75 c$_{2}$ + 0.54 \\
\hline
\end{tabular}
\end{center}
\end{table}

\newpage

\begin{table}[tb]
\caption{Deviations of Non-CCM Sightlines \label{tab:deviations_list}}
\begin{center}
\begin{tabular}{lccccc}
\hline \hline
 HD & $\delta(4.65)$ & $\delta(4.90)$ & $\delta(5.07)$ & $\delta(5.24)$ & $\delta(7.82)$ \\
\hline 
29647 & -0.58 & -0.35 & -0.18 & -0.08 & 0.31 \\
62542 & -1.49 & -1.10 & -0.75 & -0.50 & 1.12 \\
204827 & -0.65 & -0.44 & -0.29 & -0.18 & 0.43 \\
210121 & -2.63 & -2.35 & -1.76 & -1.24 & 2.33 \\
\hline
\end{tabular}
\end{center}
\end{table}

\newpage

\clearpage

\clearpage

\end{document}